\documentclass[12pt]{article}
\usepackage{amsfonts}
\usepackage{amssymb}
\usepackage{amsbsy}
\usepackage{mathrsfs}
\usepackage{amsmath}
\usepackage{epsfig}      
\usepackage{color}       

\newlength{\TZ}
\newcommand{\BEQ}{\begin{equation}}     
\newcommand{\BEA}{\begin{eqnarray}}
\newcommand{\BD}{\begin{displaymath}}
\newcommand{\EEQ}{\end{equation}}       
\newcommand{\EEA}{\end{eqnarray}}
\newcommand{\ED}{\end{displaymath}}
\newcommand{\bb}{\begin{eqnarray}}
\newcommand{\ee}{\end{eqnarray}}

\newcommand{\e}{{\rm e}}

\newcommand{\D}{{\rm d}}                
\newcommand{\II}{{\rm i}}               
\newcommand{\artanh}{{\rm artanh\,}}    
\newcommand{\erfc}{{\rm erfc\,}}        
\newcommand{\erf}{{\rm erf\,}}          

\newcommand{\wit}[1]{\widetilde{#1}}    

\renewcommand{\vec}[1]{\boldsymbol{#1}} 

\newcommand{\appsection}[2]{\setcounter{equation}{0}\setcounter{subsection}{0}
\section*{Appendix #1. #2}
\renewcommand{\theequation}{#1.\arabic{equation}}
              \renewcommand{\thesection}{#1} }

\catcode`\@=11
\def\numberbysection{\@addtoreset{equation}{section}
        \def\theequation{\thesection.\arabic{equation}}}
\numberbysection

\definecolor{gruen}{rgb}{0,0.625,0}       
\definecolor{rot}{rgb}{0.75,0,0}          
\definecolor{blau}{rgb}{0,0,0.75}         
\definecolor{casta}{rgb}{0.45,0.20,0}     
\definecolor{gelb}{rgb}{0.825,0.725,0.0}  

\begin{document}

\begin{titlepage}

\vskip 1.5 cm
\begin{center}
{\Large \bf Finite-size scaling in the ageing dynamics of the\\[0.3cm] $1D$ Glauber-Ising model\footnote{{\it In memoriam} Ralph Kenna.}} 
\end{center}

\vskip 2.0 cm
\centerline{ {\bf Malte Henkel}$^{a,b}$
}
\vskip 0.5 cm
\begin{center}
$^a$Laboratoire de Physique et Chimie Th\'eoriques (CNRS UMR 7019),\\  Universit\'e de Lorraine Nancy,
B.P. 70239, F -- 54506 Vand{\oe}uvre l\`es Nancy Cedex, France\\~\\
$^b$Centro de F\'{i}sica Te\'{o}rica e Computacional, Universidade de Lisboa, \\Campo Grande, P--1749-016 Lisboa, Portugal\\~\\
\end{center}

\begin{abstract}
Single-time and two-time correlators are computed exactly in the $1D$ Glauber-Ising model after a quench to zero temperature 
and on a periodic chain of finite length $N$, using a simple analytical continuation technique. 
Besides the general confirmation of finite-size scaling in non-equilibrium dynamics, this allows to test the scaling behaviour of the
plateau height $C_{\infty}^{(2)}$ to which the two-time auto-correlator converges, when deep into the finite-size regime. 
~\\~\\
\end{abstract}

\vfill

\end{titlepage}

\setcounter{footnote}{0}

\section{Introduction}

An important class of physical phenomena arises in the context of {\em ageing phenomena} 
\cite{Arce21,Vinc24} after a many-body system has been quenched, from 
some prescribed initial state, either onto a critical point where at least two physical phases become indistinguishable or else
into a phase co-existence region where two macroscopic physical phases coexist. In either case, the after-quench dynamics is a
slow one, which may come from the effects of the critical-point fluctuations or else from the competition between the relaxation 
towards at least two distinct physical states. Microscopically, the system separates into many (correlated or ordered) clusters 
whose mean size $\ell(t)$ is growing with time. Phenomenologically one observes, on a macroscopic scale, the
three defining properties of {\em physical ageing}, namely \cite{Stru78}
\begin{enumerate}
\item slow dynamics (relaxations are slower than might be described by simple exponentials)
\item absence of time-translation-invariance
\item dynamical scaling
\end{enumerate}
These manifest itself in typical behaviour of correlation functions, which might be thought of in terms of a coarse-grained order-parameter
$\phi=\phi(t,\vec{r})$ which depends on the time $t$ and the space coordinates $\vec{r}$. 
For example, notably in situations where the average order-parameter 
$\langle \phi(t,\vec{r})\rangle=0$, one often considers single-time
or two-time correlators (which depend on both the waiting time $s$ and the observation time $t>s$)
\BEQ
C(t;\vec{r})   = \left\langle \phi(t,\vec{r})\phi(t,\vec{0})\right\rangle  \;\; , \;\;
C(t,s;\vec{r}) = \left\langle \phi(t,\vec{r})\phi(s,\vec{0})\right\rangle
\EEQ
In this work, we restrict to {\em phase-ordering}, which occurs for a non-conserved order-parameter quenched to $T<T_c$.
Then one generically finds, for large enough times (spatial translation- and rotation-invariance are implicitly admitted)
\BEQ \label{1.2}
C(t;r)   := C(t;|\vec{r}|)   = F_C\left( 1, \frac{|\vec{r}|}{t^{1/z}}\right)            \;\; , \;\;
C(t,s;r) := C(t,s;|\vec{r}|) = F_C\left( \frac{t}{s}, \frac{|\vec{r}|}{s^{1/z}}\right)
\EEQ
which is specified here for systems where the typical domain size $\ell(t)\sim t^{1/z}$ increases algebraically for large times. 
This defines the {\em dynamical exponent} $z$.  
For a non-conserved order-parameter, and short-ranged interactions to which we shall restrict throughout, 
one has $z=2$ \cite{Bray94a,Bray94b}.\footnote{For a conserved order-parameter, 
one speaks of {\em phase separation} and $z$ takes different values \cite{Bray94a,Bray94b}. 
Long-range interactions lead to further modifications \cite{Bray94b,Chris19,Chris20,Corb19a,Corb19b,Corb24,Muel24}.}  
Up to metric scale factors, the 
form of the scaling function $F_C$ is generically expected to be universal, hence independent of most of the `details' of the underlying microscopic physics,
see \cite{Bray94a,Cugl03,Puri09,Henk10,Taeu14} for reviews.  
Knowing the form of $F_C$ is an important theoretical task and is also of practical importance since {\it a priori} 
knowledge of $F_C$ would permit to make long-time predictions
on the basis of short-time data. The expectations (\ref{1.2}) are also confirmed experimentally \cite{Maso93,Alme21}. 

\begin{figure}[tb]
\begin{center}
\includegraphics[width=.46\hsize]{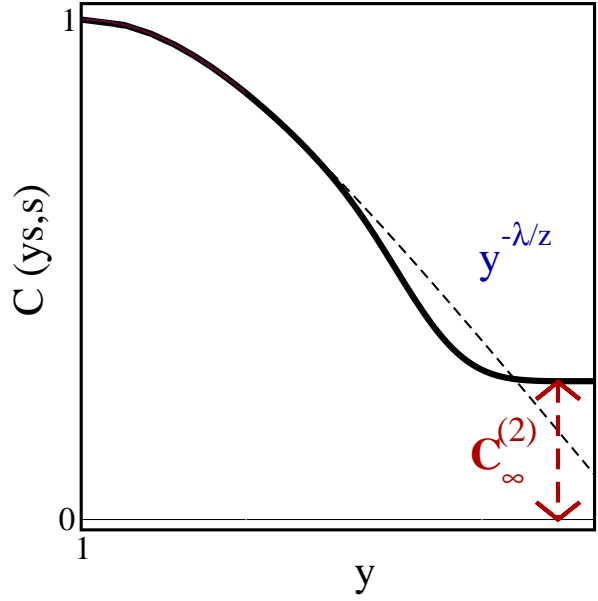} 
\end{center}
\caption[fig1]{\small{Qualitative dependence of the scaled two-time auto-correlator $C(t,s)$ on the time ratio $y=t/s$ for (i) a spatially infinite
system (dashed line) with the power-law behaviour $\sim y^{-\lambda/z}$ and (ii) in a fully finite system (full line) which converges to
a characteristic plateau $C_{\infty}^{(2)}$.} 
\label{fig1} }
\end{figure}

In practical situations, it may be difficult to achieve spatially totally homogeneous samples without any kind of interfaces and/or granular effects. 
It is therefore of interest to study situations of physical ageing in geometries which a finite extent, 
for example of a hyper-cubic form with $N^d$ sites. A simple example illustrates 
a typical kind of finite-size effect, see figure~\ref{fig1}. 
If one considers the auto-correlation function\footnote{Please distinguish carefully the two-time
auto-correlator $C(t,s)$, eq.~(\ref{1.3.a}) below, and the time-space correlator $C(t;r)$, eq.~(\ref{1.2}).} 
$C(t,s) :=C(t,s;\vec{0})$ of a phase-ordering (or phase-separating) system, 
one finds for the spatially infinite system and for sufficiently large times such that $t,s\gg \tau_{\rm micro}$ and $t-s\gg \tau_{\rm micro}$ that  
(i) a data collapse occurs and (ii) the characteristic power-law behaviour
\begin{subequations} \label{1.3.}
\begin{align} \label{1.3.a}
C(t,s) &= F_C\left(\frac{t}{s},|\vec{0}|\right) \:=\: f_C\left( \frac{t}{s}\right) \;\; , \;\; f_C(y) \stackrel{y\gg 1}{\sim} y^{-\lambda/z}
\end{align}
for large time ratios $y=t/s > 1$. Here, $\lambda$ is the {\em autocorrelation exponent} and is independent of the equilibrium critical exponents. 
A recent list of estimates of $z$ and $\lambda$ for phase-ordering is in \cite{Henk23}. 
On the other hand, in a fully finite system, even if the auto-correlator should still be close to the one of the spatially infinite system for 
$y$ not too large, there will arise deviations from (\ref{1.3.a}), see again figure~\ref{fig1}. 
Generically, in a finite system the auto-correlator should first decrease more fast as a function of $y$ than it would be the case for
the infinite system. For even larger values of $y$, the auto-correlator saturates at a plateau, of height
\begin{align} \label{1.3.b}
\lim_{y\to\infty} C(ys,s;{0};N)= C_{\infty}^{(2)}(s,N)
\end{align}
\end{subequations} 
which in principle should depend on the waiting time $s$ and the system size $N$. Qualitative discussions on this go back a long time, 
see \cite{Joh00,Kenn18}.\footnote{The systematic study of finite-size effects, 
by which we mean the consequences of the system being in a restricted spatial volume of linear size $N$, and 
the associated finite-size scaling has a long history indeed for equilibrium phase transitions \cite{Fish71,Brez82,Barb83,Bran00} 
and as well for equilibrium dynamics \cite{Suzuki77}. 
Early examples of finite-size studies in non-equilibrium systems include \cite{Krebs94a,Krebs94b,Alca94}. 
Finite-size effects in glassy dynamics are studied in \cite{Fern19,Barb21,Zamp22} and experimentally in \cite{Joh00,Kenn18,Zhai17,Zhai19}. 
They are also one possibility to artificially create spurious sub-ageing effects \cite{Chris22}.} 
Another early observation of this saturation effect occurs in the Kuramoto model of self-synchronisation \cite{Ioni14}.

The expected limit behaviour of (\ref{1.3.b}) can be understood heuristically \cite{Henk23,Wark25}. We recall the argument for quenches to $T<T_c$. 
For large times, one expects that the auto-correlator in terms of domain sizes $\ell(t)$ and $\ell(s)$ reads 
$C(t,s)\sim \left( \frac{\ell(t)}{\ell(s)}\right)^{-\lambda}$, see eq.~(\ref{1.3.a}). 
If the observation time $t$ becomes so large that the domain size $\ell(t)\sim N$ has crossed over into the saturation regime,
while the waiting time $s$ is still small enough that the infinite-system rule $\ell(s)\sim s^{1/z}$ applies (hence $\ell(s)\ll N$), one would find
\BEQ \label{1.4}
C_{\infty}^{(2)} \sim N^{-\lambda}  \mbox{\rm\small ~~~if $s$ is fixed} \;\;\; , \;\;\;
C_{\infty}^{(2)} \sim s^{\lambda/z} \mbox{\rm\small ~~~if $N$ is fixed} 
\EEQ
More formally, (\ref{1.4}) is one of the several consequences of the hypothesis of generalised time-translation-invariance \cite{Henk25}. 
Furthermore, one can write generalisations for quenches to all temperatures $T\leq T_c$, conserved and non-conserved order parameters and so on. 
One interest of (\ref{1.4}) is that it offers a new way to numerically estimate the exponents $\lambda$ and $\lambda/z$, respectively. 

The present work strives at obtaining a test of (\ref{1.4}) in the context of an exactly solvable model. 
Since there already exists an exact confirmation of (\ref{1.4}) in the spherical model for dimensions $2<d<4$ and $T<T_c$ \cite{Henk23}, 
we consider here the case of the $1D$ Glauber-Ising model,\footnote{For a short summary of the r\^ole of the Ising model in equilibrium phase transitions, 
discovered by Cagniard de la Tour about 200 years ago, see e.g. \cite{Berc09a,Berc09b} and refs. therein.} 
quenched to temperature $T=0$ from a fully disordered initial 
state\footnote{Initial correlations are irrelevant at large times in the $1D$ Glauber-Ising model \cite{Henk04}.} 
and whose single- and two-time correlators obey the scaling form (\ref{1.2}). 
Tests of (\ref{1.4}) in the $2D$ Glauber-Ising model quenched to $T<T_c$ will be presented elsewhere \cite{Wark25}. 
We shall be interested in deriving the full size-dependent single-time and two-time spin-spin correlators 
which allows at the end to perform an explicit test of (\ref{1.4}).

This work is organised as follows. In section~2, we shall introduce the analytic continuation technique 
and confirm that it reproduces the known exact results. 
In section~3, we shall use it to compute the finite-size effects in the ageing dynamics and finally confirm (\ref{1.4}). Section~4 gives our
conclusions. Technical details of the exact solution are given in five appendices. 

\section{Critical relaxations in infinite-size systems}

\subsection{The $1D$ Glauber-Ising model} 

The nearest-neighbour Ising model on a chain $\Lambda \subset \mathbb{Z}$ is defined through the hamiltonian
\BEQ
{\cal H} = - \sum_{n\in\Lambda} \sigma_{n} \sigma_{n+1}
\EEQ
for the Ising spins $\sigma_n = \pm 1$ and the exchange coupling was normalised to unity. 
In a heat-bath formulation, at temperature $T$, 
at each time step $\Delta t$ a randomly chosen site $n\in\Lambda$ is updated according to Glauber dynamics \cite{Glau63} with the 
rates \cite{Godr00a} 
\BEQ \label{2.2}
\sigma_n(t) \mapsto \pm 1 \mbox{\rm\small ~~~with the probability $\frac{1}{2}\left( 1 \pm \tanh \frac{\sigma_{n-1}(t) + \sigma_{n+1}(t)}{T}\right)$}
\EEQ
On a discrete chain, the single-time correlator is $C_n(t) := \left\langle \sigma_{n}(t)\sigma_{0}(t)\right\rangle$
where the average is over the thermal histories defined by eq.~(\ref{2.2}). The correlator obeys the equation of motion \cite{Glau63,Godr00a,Lipp00,Maye04}
\BEQ \label{2.3}
\partial_t C_n(t) = -2 C_n(t) + \gamma \bigl( C_{n-1}(t) + C_{n+1}(t) \bigr) \mbox{\rm\small ~~~when $n\ne 0$~~} \;\; , \;\; C_0(t) = 1 
\EEQ
with the abbreviation $\gamma=\tanh(2/T)$ (such that $0\leq \gamma\leq 1$) and the microscopic rate constant was normalised to unity. 
An initial condition must still be given; for an initially fully disordered system one has 
$C_n(0)=\delta_{n,0}$ \cite{Glau63,Godr00a}. In this, spatial translation-invariance is implicitly admitted and we shall do so throughout. 
We shall give later the 
periodicity conditions for systems on a lattice of finite size $N$. 
Similarly, with the time difference $\tau=t-s$, the two-time correlator is defined as $C_{n}(\tau,s) := \left\langle \sigma_{n}(t)\sigma_{0}(s)\right\rangle
= \left\langle \sigma_{n}(\tau+s)\sigma_{0}(s)\right\rangle$ and obeys the equation of motion \cite{Glau63,Godr00a,Lipp00,Maye04}
\BEQ \label{2.4}
\partial_{\tau} C_n(\tau,s) = -C_n(\tau,s) + \frac{\gamma}{2} \bigl( C_{n-1}(\tau,s) + C_{n+1}(\tau,s) \bigr) \;\; , \;\; 
C_n(0,s) = C_n(s)
\EEQ
Herein, the single-time correlator $C_n(s)$ serves as initial value. 
Solving eqs.~(\ref{2.3},\ref{2.4}) constitutes the mathematical problem for the determination of the single-time and two-time correlators.  

\subsection{The discrete case} 

There are many well-known ways \cite{Glau63,Feld71,Droz89,Bray97,Godr00a,Lipp00,Maye03,Maye04,Maye05,Henk04,Alie09,Krap10,Verl11,Godr22} 
to solve eqs.~(\ref{2.3},\ref{2.4}), 
using for example generating functions \cite{Glau63}, free fermions \cite{Feld71}, a scaling ansatz \cite{Bray97,Krap10}, Laplace transforms \cite{Godr00a}, 
grassmannian variables \cite{Alie09} or systems
of equations of motion \cite{Lipp00,Maye03,Maye04,Maye05,Henk04,Verl11}. 
Here, we shall adopt a method which easily generalises to finite systems as well.

We begin with the single-time correlator.\footnote{For a disordered initial condition $\langle\sigma_n(0)\rangle=0$, 
the site-dependent magnetisation $\langle\sigma_n(t)\rangle=0$ for all times.} 
In principle, one wishes to decouple the equations of motion (\ref{2.3}) 
by a Fourier transform, but because of the boundary condition $C_0(t)=1$ this is not straightforward. 
Rather, we observe first that the single-time correlator
$C_n(t)=\left\langle\sigma_n(t)\sigma_0(t)\right\rangle=\left\langle\sigma_0(t)\sigma_n(t)\right\rangle=\left\langle\sigma_{|n|}(t)\sigma_0(t)\right\rangle$ 
is even in $n$ and does
not depend on the sign of $n$. It is enough to restrict the physical interpretation to positive values of $n>0$. In particular, the equation of motion
(\ref{2.3}) is needed for $n>0$ only. 
Therefore, we consider that $C_{-n}(t)$ can be thought of as an entity devoid of physical significance. 
It can therefore be used for purely calculational, mathematical purposes. We shall {\em define} $C_{-n}(t)$ in such a way that the equation of motion
(\ref{2.3}) holds for all values of $n\in\mathbb{Z}$. To do so, we write down the ansatz
\BEQ \label{2.5}
C_{-n}(t) = \alpha_n - C_{n}(t) \;\; ; \;\; n\geq 0
\EEQ
and try to choose $\alpha_n$ such that the boundary condition $C_0(t)=1$ and the equation of motion (\ref{2.3}) become valid for all $n\in\mathbb{Z}$. 
This kind of analytical continuation will be used several times 
below.\footnote{It had already been used in the exact treatment of $1D$ coagulation-diffusion processes \cite{Durang10,Durang11,Fort14}, to find
the single-time and two-time correlation and response functions.}  
It is easy to check that this is indeed possible if the $\alpha_n$ obey the (time-independent) recursion relation
\BEQ \label{2.6}
\alpha_{n+1} = \frac{2}{\gamma} \alpha_n - \alpha_{n-1} \;\; , \;\; \alpha_0 = 2 \;\; , \;\; \alpha_1 = \frac{2}{\gamma}
\EEQ

\noindent
{\bf Lemma:} {\it The solution of the recursion relation (\ref{2.6}) is for all $n\geq 0$}
\BEQ \label{2.7} 
\alpha_n = \left( \frac{\gamma}{1 - \sqrt{1-\gamma^2\,}\,}\right)^n + \left( \frac{\gamma}{1 + \sqrt{1-\gamma^2\,}\,}\right)^n
\EEQ

\noindent {\bf Proof:} This is easily checked by induction. For $n=0$ and $n=1$, the initial values in (\ref{2.6}) are immediately reproduced. 
Then it is straightforward to check that (\ref{2.7}) obeys indeed the recurrence relation (\ref{2.6}). \hfill q.e.d. \\

It then follows that one has for all $n\in\mathbb{Z}$ the linear equation of motion
\BEQ \label{2.8}
\partial_t C_n(t) = -2 C_n(t) + \gamma \bigl( C_{n-1}(t) + C_{n+1}(t) \bigr) 
\EEQ
but now in the absence of any boundary condition. This is the goal we wanted to achieve. The solution of eq.~(\ref{2.8}) is described in appendix~A. 
It follows that the physical correlator is
\BEQ \label{2.9}
C_n(t) =  \left\langle\sigma_{|n|}(t)\sigma_{0}(t)\right\rangle = \eta_{-}^{|n|}  
+\sum_{m=1}^{\infty} \left( C_m(0)-\eta_{-}^m\right) e^{-2t}\left[ I_{|n|-m}(2\gamma t) - I_{|n|+m}(2\gamma t)\right] 
\EEQ
where $I_n(t)$ is a modified Bessel function \cite{Abra65} and 
\BEQ \label{2.10}
C_{n,{\rm eq}} = \eta_{-}^{|n|} = \left( \frac{1 - \sqrt{1-\gamma^2\,}\,}{\gamma} \right)^{|n|} 
\EEQ
is the equilibrium correlator $C_{n,{\rm eq}}=\left\langle \sigma_{|n|}\sigma_{0}\right\rangle_{\rm eq}$. 
This is indeed nothing else than Glauber's time-honoured result \cite[(eq. (63)]{Glau63}. 
For any $0\leq \gamma <1$, it shows the rapid relaxation towards the equilibrium correlator (\ref{2.10}) 
on a finite time-scale $\tau_{\rm rel}^{-1}=2(1-\gamma)$. 
Formally, $\tau_{\rm rel}$ diverges if $\gamma\to 1$ (or the temperature $T\to 0$) 
and ageing is only possible in this latter case, see section~1. 

This whole calculation only has the purpose of showing that the analytical continuation technique used 
here does reproduce the well-known results in the literature 
\cite{Glau63,Godr00a,Maye03,Maye04,Henk04}. Having verified this, 
we can now proceed towards the derivation of new results, notably on the critical dynamics, with $\gamma=1$. 

For the two-time correlator, it is straightforward to solve (\ref{2.4}) 
by a Fourier transform and then insert the single-time correlator (\ref{2.9}). Since this leads
to lengthy expressions \cite{Maye03}, which are not needed below, we do not carry this out explicitly. 

\subsection{The continuum limit} 

{\bf 1.} Having seen that ageing can only occur for $\gamma=1$, we restrict to this case from now on. The calculations become more short in the 
continuum limit,\footnote{This is valid in the scaling limit of (\ref{2.3}) where $t\to\infty$, $n\to\infty$ but $n^2/t$ is kept fixed. This will always
be taken in what follows, where $t$ and $x$ are re-scaled time and space coordinates.} where
for the single-time correlator $C(t;x)=\left\langle \sigma(t,x) \sigma(t,0)\right\rangle$ we have from (\ref{2.3}) the equation of motion
\BEQ \label{2.11}
\partial_t C(t;x) = \partial_x^2 C(t;x) \;\;\; , \;\;\; C(t;0) = 1
\EEQ
where the diffusion constant was scaled to unity. Since the physical correlator $C(t;x)$ 
is even in $x$, we can restrict the problem (\ref{2.11}) to the half-line
$x\geq 0$ and use the function $C(t;-x)$ for computational purposes. The analytic continuation (\ref{2.5}) then simplifies to
\BEQ \label{2.12}
C(t;-x) = 2 - C(t;x) 
\EEQ
for $x\geq 0$. In appendix~B, we show that (with the complementary error function $\erfc(x)$ \cite{Abra65}) 
\BEQ \label{2.13}
C(t;x) = \erfc\left(\frac{|x|}{2 t^{1/2}}\right) 
         + \frac{e^{-x^2/4t}}{\sqrt{\pi\,t\,}} \int_0^{\infty} \!\D x'\: C(0;x'\,)\, e^{-x'^2/4t} \sinh\left(\frac{|x|\,x'}{2t}\right)
\EEQ
still for any initial correlations $C(0;x)$. For an initially fully disordered lattice $C(0;x)=0$ for $x>0$ and the second term in (\ref{2.13}) vanishes, 
as stated countless times in the literature, see e.g. \cite{Glau63,Bray94a,Bray97,Godr00a,Lipp00,Maye04,Krap10} and references therein. 
In order to clarify further its importance, consider as an example 
$C(0;x)= \left( 1 + x^2\right)^{-\aleph/2}$ with a positive exponent $\aleph>0$. For large times, $C(t;x) = \mathfrak{T}_1 + \mathfrak{T}_2$ 
where the contribution of the second term $\mathfrak{T}_2$ will be of the order
\BD
\mathfrak{T}_2 \sim e^{-x^2/4t} \int_0^{\infty} \!\D x'\: t^{-\aleph/2}\,e^{-x'^2/4} \sinh\left(\frac{|x| x'}{2 t^{1/2}}\right) \sim t^{-\aleph/2} \to 0 
\ED
when $t\to\infty$ but where $x^2/t$ is kept fixed. 
This is much more small than the scale-invariant and finite first term $\mathfrak{T}_1=\erfc\bigl(|x|/2 t^{1/2}\bigr)$. 
We conclude that in the $1D$ Glauber-Ising model at $T=0$, initially decaying correlations are always irrelevant for $t\to\infty$. 
This merely confirms long-standing results for the two-time correlator in this model \cite{Henk04}. 
Consequently, only the universal first term in (\ref{2.13}) 
is important in the long-time scaling limit and the initial-condition-dependent second term can be discarded. 
The leading term in (\ref{2.13}) might have been found more readily by inserting the scaling ansatz (\ref{1.2}) into (\ref{2.11}) \cite{Bray97}. 

\begin{figure}[tb]
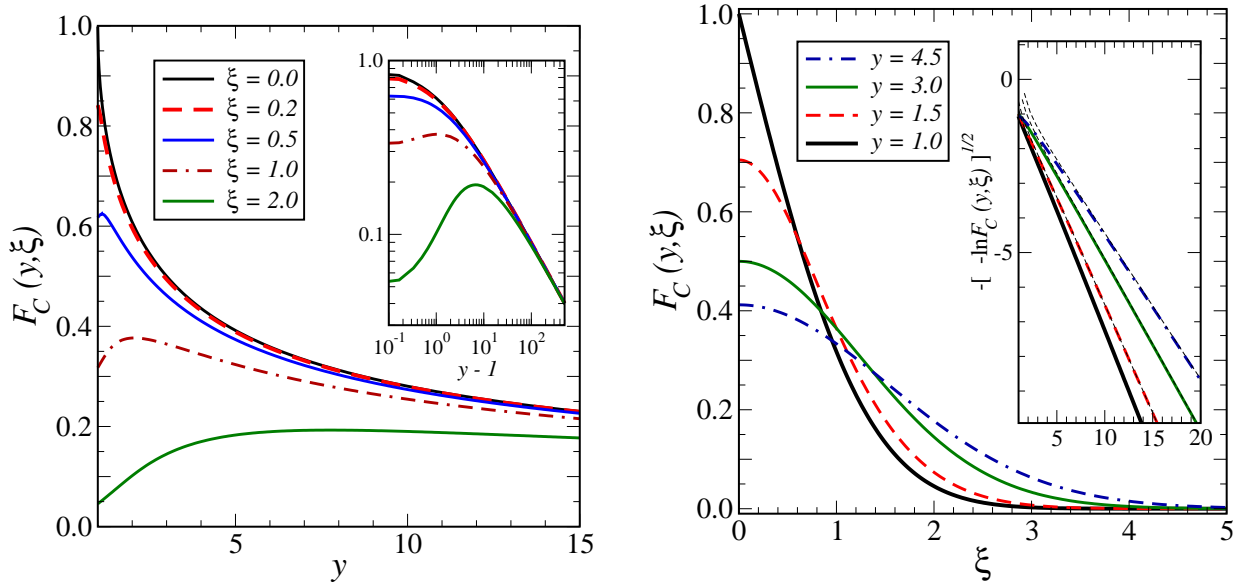

\begin{center}
\includegraphics[width=.46\hsize]{Glauber-ETF_Corr-y3.eps} ~~~ \includegraphics[width=.46\hsize]{Glauber-ETF_Corr-xi3.eps}
\end{center}
\caption[fig2]{\small{Properties of the scaled two-time correlator (\ref{1.2}) in the $1D$ Glauber-Ising model.\\
\underline{Left panel:} universal decay of the correlator $F_C(y,\xi)$ for large $y$, with $\xi=[0.0,0.2,0.5,1.0,2.0]$ from top to bottom. 
The inset shows the expected universal power-law decay (\ref{2.19}) for large values of $y$. \\
\underline{Right panel:} decay of the correlator $F_C(y,\xi)$  as a function of $\xi$ for for 
$y=[1.0,1.5,3.0,4.5]$ from bottom to top on the right of the figure. 
The inset highlights the expected gaussian decay for large $\xi$ and the dashed lines indicate the leading decay behaviour (\ref{2.20}).} 
\label{fig2} }
\end{figure}

Using the notation of (\ref{1.2}), and for an initially fully disordered lattice, 
the scaled correlator $C(t;x)=F_C(1,|x|/2\sqrt{t\,})$ is shown in the right panel of figure~\ref{fig2} 
as the full black curve labelled $y=1.0$, as a function of the scaling variable
$\xi^2 = x^2/2 t$. For small arguments, one observes a sharp peak at $\xi=0$. 
This indicates that the interfaces between domains are sharp, as expected for kinetic
Ising models and in agreement with Porod's law \cite{Bray94a,Henk10}. 

{\bf 2.} In appendix~B, we also compute the characteristic length scale $\ell(t)\sim t^{1/z}$ from the second moment of $C(t;x)$ and find
\BEQ \label{2.14}
\ell^2(t) = \frac{\int_0^{\infty}\! \D x\: x^2\, C(t;x)}{\int_0^{\infty}\! \D x\:  C(t;x)} = \frac{4}{3}\, t
\EEQ
which is exact for a fully disordered initial state. Eq.~(\ref{2.14}) confirms the expected \cite{Bray94b} dynamical exponent $z=2$. 
One might use it in (\ref{2.13}) to achieve a data collapse $C(t;x) = \erfc\bigl( |x|/(\sqrt{3\,}\,\ell(t))\bigr)$, up to irrelevant terms.  
 
{\bf 3.} We now turn to the two-time correlator
\BEQ
C(\tau,s;x) := \left\langle \sigma(\tau+s,x) \sigma(s,0)\right\rangle
\EEQ
For the zero-temperature case $\gamma=1$, we have from (\ref{2.4}) the equation of motion
\BEQ \label{2.16}
\partial_{\tau} C(\tau,s;x) = \frac{1}{2} \partial_x^2 C(\tau,s;x) \;\; , \;\;
C(0,s;x) = C(s;x) = \erfc\left(\frac{|x|}{2 s^{1/2}}\right)
\EEQ
where in the initial condition, we merely retained from (\ref{2.13}) the most relevant term for large times or use a totally disordered initial state. 
In what follows, we shall use the scaling variables
\BEQ \label{skal-var}
y = \frac{t}{s} >1 \;\; , \;\; \xi^2 = \frac{x^2}{2s}
\EEQ
for the ratio $y$ of the two times and the re-scaled length $\xi$. Then the time difference $\tau=t-s=s(y-1)$. 
{}From now on, we shall always work in the long-time limit $s\to\infty$ with $y$ and $\xi$ being kept fixed. 
In appendix~B, we show that (\ref{2.16}) is solved by 
\begin{subequations} \label{2.17}
\begin{align}
C(\tau,s;x) &= 1 - \frac{1}{\pi} e^{-x^2/(2\tau)} \left(\frac{2\tau}{s}\right)^{1/2} 
\Psi_1\left(1,\frac{1}{2};\frac{3}{2},\frac{1}{2};-\frac{\tau}{2s},\frac{x^2}{2\tau}\right) \label{2.17a} \\
&= 1 - \frac{2}{\pi} e^{-\xi^2/(y-1)} \left( \frac{y-1}{2}\right)^{1/2} 
\Psi_1\left(1,\frac{1}{2};\frac{3}{2},\frac{1}{2};-\frac{y-1}{2},\frac{\xi^2}{y-1}\right) 
\label{2.17b} 
\end{align}
\end{subequations}
where $\Psi_1$ is a Humbert confluent hyper-geometric function of two variables \cite{Sriv85,Prud3}. 
In the scaled expression (\ref{2.17b}), we observe the independence from the waiting time $s$.
In what follows, with a slight change of notation with respect to (\ref{1.2}), 
we shall write more shortly $F_C(y,\xi):=C(\tau,s;x)$ in terms of the scaling variables $y$ and $\xi$, defined above in (\ref{skal-var}).  
To the best of our knowledge, previous efforts concentrated on the two-time auto-correlator 
$C(\tau,s;0)$.\footnote{When considering the scaling influence of the temperature
through $\gamma$, the two-time {\em auto}-correlator can be expressed via another Humbert function $\Phi_1$ \cite{Godr00a}. 
Taking into account all three scaling variables
will likely involve a three-argument Lauricella/Horn hyper-geometric series.} 

To recover the known two-time auto-correlator $C(ys,s)=F_C(y,0)$, we now recall that 
$\Psi_1(a,b;c,c';x,0)={}_2F_1(a,b;c;x)$ reduces to Gau{\ss}' hyper-geometric function 
which in turn is related to elementary functions in our example \cite[(7.3.1.123)]{Prud3}, \cite[(7.3.2.83)]{Prud3}. This gives 
\BEQ \label{autoC}
F_C(y,0) = 1 -\frac{\sqrt{2\,}}{\pi}\left(\frac{\tau}{s}\right)^{\frac{1}{2}} \left(-\frac{\tau}{2s}\right)^{-\frac{1}{2}} \artanh\sqrt{-\frac{\tau}{2s}\,} 
= \frac{2}{\pi} \arctan\sqrt{\frac{2s}{\tau}\,} =  \frac{2}{\pi} \arctan\sqrt{\frac{2}{y-1}\,}
\EEQ
where also \cite[(4.4.42)]{Abra65} was used. The last two expressions reproduce the well-known 
\cite[eq.~(4.24)]{Godr00a}, \cite[eq.~(17)]{Lipp00}, as expected. 
Especially, for $y\gg 1$, one has
$F_C(y,0)\simeq \frac{2\sqrt{2\,}}{\pi} y^{-1/2}$ and by comparison with the standard expected asymptotics 
(\ref{1.3.a}) one reads off $\frac{\lambda}{z}=\frac{1}{2}$ or
since $z=2$ \cite{Bray94b}, the auto-correlation exponent $\lambda=1$ \cite{Godr00a,Lipp00,Henk04}, 
as it should be. In the left panel of figure~\ref{fig2}, the
auto-correlator is the full black line labelled $\xi=0.0$. 

Several mathematical identities, listed in appendix~E, can be used to check the expressions (\ref{2.17}) and to extract a few consequences. 
First, the limit relation given in Lemma~E.1 allows to confirm
that in the limit $\tau\to 0$ or $y\to 1$, the expressions (\ref{2.17}) 
indeed reduce to the single-time correlator (\ref{2.13}), but of course without the irrelevant and sub-dominant
terms coming from the initial conditions. This is illustrated in the left panel of figure~\ref{fig2}, 
where the curves become more close to the auto-correlator when $\xi\to 0$. 
Second, the asymptotic relation given in Lemma~E.2, especially the independence of the two leading terms on the 
second argument, describes what happens when $y=t/s$ is getting large. In figure~2, the left panel shows that the asymptotic behaviour is
\BEQ \label{2.19}
F_C(y,\xi) \stackrel{y\gg 1}{\simeq} \frac{\sqrt{8\,}}{\pi} \frac{1}{y^{1/2}} \left( 1 + {\rm O}(y^{-1/2}) \right)
\EEQ
and that this leading behaviour is universal in the sense that is independent of the scaled distance $\xi^2=\frac{x^2}{2s}$. 
This is reproduced from (\ref{E.4}) since the leading term cancels but the next order does give (\ref{2.19}). 
Third, one may also investigate the
leading behaviour of the correlator when $\xi$ is getting large but $y$ is kept fixed. This is shown in the right panel of figure~2. 
While the main figure merely displays quite
a rapid decay with increasing $\xi$, the inset shows that the plot of the square root of the logarithm of $F_C$, 
namely $-\sqrt{ - \ln F_C(y,\xi)\,}$ over against $\xi$, produces an almost straight line as a
function of $\xi$, but with a $y$-dependent slope. 
Lemma~E.3 gives the mathematical justification of this observation and, again, the second term in (\ref{E.5}) 
establishes that for $\xi\gg 1$
\BEQ \label{2.20}
F_C(y,\xi) \stackrel{\xi\gg 1}{\simeq} \frac{1}{\sqrt{2\pi\,}} \frac{y+1}{\xi} \exp\left(-\frac{\xi^2}{y+1}\right)
\EEQ
The dashed lines in the inset are these asymptotic predictions, for several values of $y\geq 1$. 
Again, if one takes the limit $y\to 1$ in (\ref{2.20}), one recovers the
leading large-distance asymptotics of the single-time correlator $\erfc(\xi/\sqrt{2\,})$, as it should be. 

\section{Critical relaxations in finite-size systems}

Having seen for the infinite system that our technique of analytical continuation, 
via (\ref{2.5}) or (\ref{2.12}), reproduces the known results, we shall now
apply it to extract the finite-size scaling properties of the ageing dynamics.

\begin{figure}[tb]
\begin{center}
\includegraphics[width=.25\hsize]{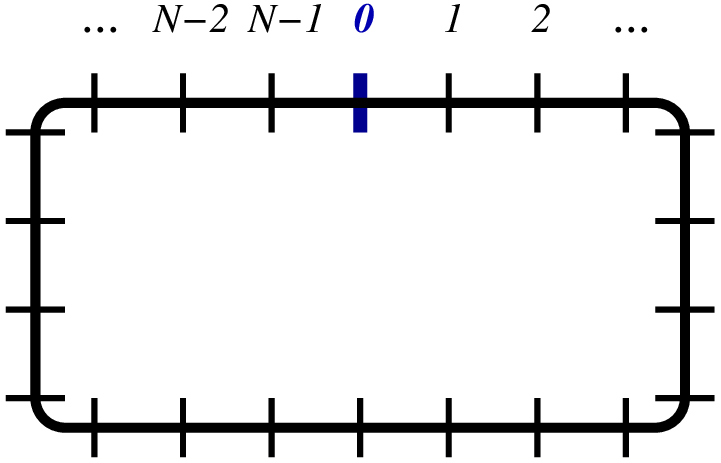}
\end{center}
\caption[fig3]{\small{Periodic ring with $N$ sites. Starting from an arbitrary site labelled $0$, 
the property $C(t;x)=C(t;N-x)$ becomes intuitive for $x\geq 1$.}
\label{fig3} }
\end{figure}

{\bf 1.} Consider the Glauber-Ising model on a periodic ring with $N$ sites, see figure~\ref{fig3}. 
In the continuum limit, the equation of motion for the single-time correlator is now
\BEQ \label{3.1}
\partial_t C(t;x) = \partial_x^2 C(t;x) \;\; , \;\; C(t;0) = C(t;N) = 1
\EEQ
The first of these boundary conditions will be treated as before by analytic continuation. 
For the second one, in the case of periodic boundary conditions, we expect to have
\BEQ \label{3.2}
C(t;x) = C(t;N-x) \;\; , \;\; C(t;-x) = 2 - C(t;x)
\EEQ
The first of these is an inversion relation suggestive from figure~\ref{fig3}. 
Both together analytically continue $C(t;x)$ from the physical region $x\in[0,N]$ towards $x\in\mathbb{R}$, via (\ref{3.2}). 
Together, they imply the periodicity relation 
\BEA
C\bigl(t;2N+x\bigr) &=& C\bigl(t;N - (-N-x) \bigr) \:=\: C\bigl(t;-N-x\bigr) \nonumber \\
&=& 2 - C\bigl(t;N+x\bigr) \hspace{0.82cm}\:=\: 2 - C\bigl(t;N-(-x)\bigr) \:=\: 2 - C\bigl(t;-x\bigr) \nonumber \\
&=& C\bigl(t;x\bigr) \label{3.3}
\EEA
where in the first line, we used the inversion condition (\ref{3.2}). 
In the second line, we applied the second continuation condition (\ref{3.2}) and then the inversion once more. 
Finally, a last application of the second condition (\ref{3.2}) brings us to the end result in the third line.

In consequence, while the physical correlator is obtained for values $0\leq x \leq N$, 
the analytically continued function $C(t;x)$ is a function of period $2N$ in $x$,
but only half of it has a physical meaning. We therefore have the Fourier representation
\BEQ \label{3.4}
C(t;x) = \sum_{k=-\infty}^{\infty} \wit{C}(t;k)\,e^{\II\pi k \frac{x}{N}} \;\; , \;\;
\wit{C}(t;k) = \frac{1}{2N} \int_{-N}^{N} \!\D x\: C(t;x) e^{-\II\pi k \frac{x}{N}}
\EEQ
Because of the periodicity condition (\ref{3.3}), we have especially
\BEQ \label{3.5}
C(t;N) = C(t;-N) \;\; , \;\; \partial_x C(t;N) = \partial_x C(t;-N)
\EEQ
which are of course compatible with the required boundary conditions in (\ref{3.1}). 

\begin{figure}[tb]
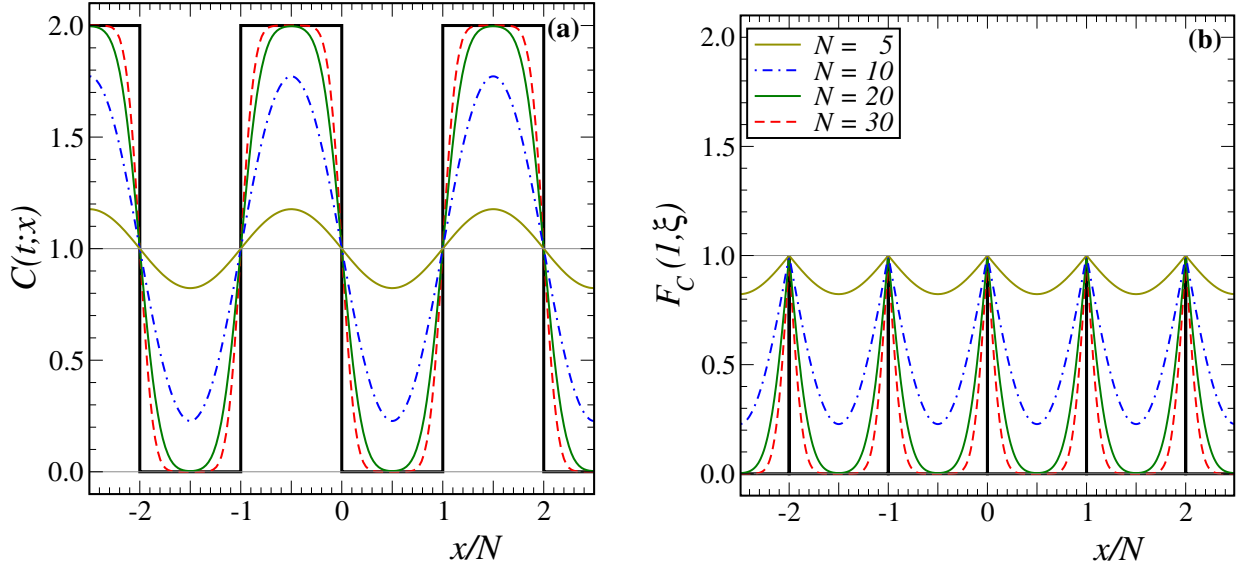

\begin{center}
\includegraphics[width=.46\hsize]{Glauber-ETF_correlateur1_anal2.eps} ~~~ \includegraphics[width=.46\hsize]{Glauber-ETF_correlateur1_phys.eps}
\end{center}
\caption[fig4]{\small{\bf (a)} Analytically continued function $C(t;x)$ as computed in appendix~C, 
for $t=5$ and $N=[5,10,20,30]$ from top to bottom, in the interval $0\leq x\leq N$. It satisfies
the periodicity conditions (\ref{3.2}). The full black line gives the initial function $C(0;x)$, for a completely disordered initial lattice. 
The thin horizontal lines indicate the values $C(t;x)=0$ and $C(t;x)=1$, respectively. \\
{\bf (b)} Physical scaling function $F_C(1,\xi)$ of eq.~(\ref{1.2}), for the same values of $t$ and $N$. 
The full black line corresponds to a completely disordered initial state. 
\label{fig4} }
\end{figure}

In this way, the boundary conditions are taken into account such that the analytically continued function 
$C(t;x)$ is a periodic solution of period $2N$, of the simple
diffusion equation $\partial_t C(t;x) = \partial_x^2 C(t;x)$. 
In appendix~C we show that for $|x|\leq N$, the physical correlator can be expressed in terms of a Jacobi theta function $\vartheta_3(z,q)$ 
\BEA
C(t;x) 
&=& \frac{1}{2N} \int_{0}^N \!\D x'\: \left\{ 2\vartheta_3\left(\frac{\pi}{2} \frac{|x|+x'}{N},e^{-\pi^2 t/N^2}\right) \right. \nonumber \\
& & \left. +C(0;x'\,) \left[ \vartheta_3\left(\frac{\pi}{2} \frac{|x|-x'}{N},e^{-\pi^2 t/N^2}\right) 
                           - \vartheta_3\left(\frac{\pi}{2} \frac{|x|+x'}{N},e^{-\pi^2 t/N^2}\right) \right] \right\}
\EEA
and that the required periodicity properties (\ref{3.2}) are indeed satisfied, if they only hold for the initial correlator $C(0;x)$. 
The term depending on the initial conditions $C(0;x)$ should be irrelevant. Then, or for a fully disordered initial state, 
one may alternatively re-write the physical single-time correlator (\ref{1.2}) as
\BEQ \label{3.7}
C(t;x) = F_C\left(1,\frac{|x|}{2 \sqrt{t\,}}\right) 
= \int_{0}^{1} \!\!\D u\: \vartheta_{3}\left(\frac{\pi}{2} u + \frac{\pi}{2}\frac{|x|}{N},e^{-\pi^2 t/N^2}\right) 
= 1 - \frac{2}{\pi}\int_{0}^{|x|/N} \!\!\!\D v\: \vartheta_2\left( \pi v,e^{-4\pi^2 t/N^2}\right) 
\EEQ
where $\vartheta_{2,3}$ are distinct Jacobi theta functions \cite{Abra65}. The expressions (\ref{3.7}) 
give finite-size scaling expressions for the single-time correlator, in terms
of the finite-size scaling variables $x/N$ and $t/N^2$. 

In figure~\ref{fig4}a, we show the analytically continued function $C(t;x)$ as computed in appendix~C. 
By construction, for $t>0$ it is an analytic function of $x$ and periodic with period $2N$,
but only in the interval $0\leq x\leq N$ it has a physical meaning. 
That function is quite distinct from the physical scaling function $F_C(1,\xi)$ 
of the single-time correlator (\ref{1.2}), with the scaling variables (\ref{skal-var}), 
which is shown in figure~\ref{fig4}b on the same scale. 
It becomes periodic in $x$ with period $N$, as expected intuitively for a periodic lattice of $N$ sites, but its derivative has jumps at $x=n N$ with 
$n\in\mathbb{Z}$. For $N\to\infty$, it converges towards the infinite-size correlator, but at $|x|=\frac{N}{2}$ 
it has a minimum value which converges exponentially to zero as $N\to\infty$. 
For a fully disordered initial state, the correlator is non-vanishing only at $x=n N$. 

{\bf 2.} A finite-size generalisation of the second moment can be given as follows 
\BEA
\ell^2(t) &=& \frac{\int_0^N \!\D x\: 8\bigl(\frac{N}{\pi} \sin\frac{\pi x}{N}\bigr)^2 C(t;x)}{\int_0^N \!\D x\:  C(t;x)}
= \frac{2 N^2}{\pi^2} \left( 1 - \frac{1}{2\pi} 
\frac{\int_0^{1} \!\D w\: \sin(2\pi w) \vartheta_3\bigl(\frac{\pi}{2}w,q\bigr)}{\int_0^{1} \!\D w\: w\, \vartheta_3\bigl(\frac{\pi}{2}w,q\bigr)}\right)
\nonumber \\
&=& \left\{ \begin{array}{ll} \frac{2}{\pi^2} N^2 & \mbox{\rm ~~~ if $t\gg N^2$}\\[0.2cm]
                            \frac{4}{3}\, t     &\mbox{\rm ~~~ if $t\ll N^2$}
            \end{array} \right.
\label{3.8}
\EEA
with the only variable $q=e^{-\pi\bigl(\pi t/N^2\bigr)}$, see appendix~C for the details. 
This reproduces the infinite-lattice result (\ref{2.14}) and has the expected finite-size scaling form $\ell^2(t)=N^2 F_{\ell}\bigl(t N^{-2}\bigr)$. 
Deep into the finite-size saturation regime, one has $\ell(t)\approx 0.45 N$.

{\bf 3.} The critical two-time correlator $C(\tau,s;x)$ satisfies the equation of motion
\BEQ \label{3.9}
\partial_{\tau} C(\tau,s;x) = \frac{1}{2}\partial_x^2 C(\tau,s;x) \;\; , \;\; C(0,s;x) = C(s;x)
\EEQ
with the single-time correlator, restricted to the relevant term, from (\ref{3.7}). In appendix~D, we show that
\BEQ \label{3.10}
C(\tau,s;x) = \frac{1}{2} \!\iint_0^1 \!\D v \D u\: \vartheta_3\left(\frac{\pi}{2}(v+u),e^{-\frac{\pi^2}{N^2}s}\right) 
\left[ \vartheta_3\left(\frac{\pi}{2}\frac{|x|}{N}-\frac{\pi}{2}v,e^{-\frac{\pi^2}{N^2}\frac{\tau}{2}}\right)  
     + \vartheta_3\left(\frac{\pi}{2}\frac{|x|}{N}+\frac{\pi}{2}v,e^{-\frac{\pi^2}{N^2}\frac{\tau}{2}}\right)  \right]
\EEQ
and hence the auto-correlator becomes
\BEQ \label{3.11} 
C(\tau,s;0) = \int_0^1 \!\D v\!\int_0^1 \!\D u\: \vartheta_3\left(\frac{\pi}{2}(v+u),e^{-\frac{\pi^2}{N^2}s}\right) 
\vartheta_3\left(\frac{\pi}{2}v,e^{-\frac{y-1}{2}\frac{\pi^2}{N^2}s}\right)  
\EEQ
which only depends on the finite-size scaling variables $s/N^2$ and $\tau/N^2=\frac{s}{N^2}(y-1)$.

\begin{figure}[tb]
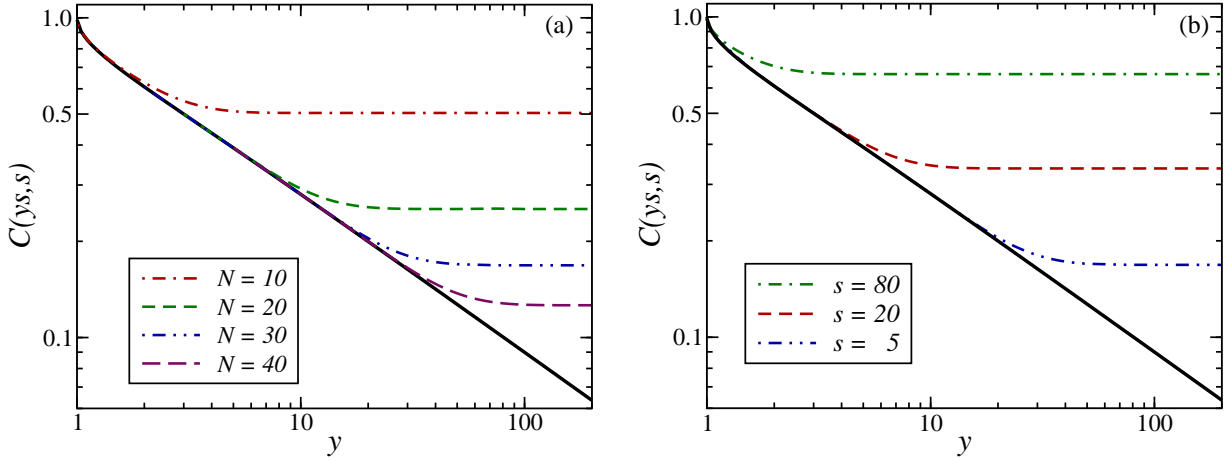

\begin{center}
\includegraphics[width=.46\hsize]{Glauber-ETF_autoC-N.eps} ~~ \includegraphics[width=.46\hsize]{Glauber-ETF_autoC-s.eps} 
\end{center}
\caption[fig5]{\small Two-time scaled auto-correlator $C(ys,s)=F_C(y,0)$ in the $1D$ Glauber-Ising model quenched to $T=0$, 
as a function of $y=t/s$ for {\bf (a)} finite systems of sizes $N=[10,20,30,40]$ from top to bottom and for
a waiting time $s=5$ and {\bf (b)} the waiting times $s=[5,20,80]$ from bottom to top and the finite size $N=30$. 
The full black line is the scaled infinite-size auto-correlator (\ref{autoC}). 
\label{fig5} }
\end{figure}

In figure~\ref{fig5}a, the scaled two-time auto-correlator $C(ys,s)=F_C(y,0)$ given in (\ref{3.11}) is shown for several values of $N$ and $s=5$ fixed. 
In analogy with the generic expectation formulated in
section~1, we find a cross-over to a plateau when $y\gg 1$ and whose height $C_{\infty}^{(2)}$ is decreasing when $N$ increases and for fixed $s$. 
However, in the $1D$ Glauber-Ising model, the cross-over towards to the plateau occurs directly and
we do not see first a more rapid decay, as it occurs in the spherical model in $2<d<4$ dimensions and $T<T_c$ \cite{Henk23} 
or in the $2D$ Glauber-Ising model at $T<T_c$ \cite{Wark25}, see figure~\ref{fig1}. 
{}From the heuristics of section~1, this plateau should be found when
the observation time $t\gg N^2$ is deeply into the finite-size regime but the waiting time $s$ not, thus $s\ll N^2$. 
When we apply this to the auto-correlator (\ref{3.11}) and let $y\gg 1$, 
the second theta function should converge rapidly towards unity, whereas the evaluation of the first one
requires to take many terms of the defining series \cite{Abra65} into account. 
This is treated by using the modular identity (\ref{C.9}) such that the auto-correlator becomes in the plateau region 
\BEA
C(\tau,s;0) &\stackrel{y\gg 1}{\simeq}&
\int_0^1\!\D v\int_0^1\!\D u\: \frac{1}{\sqrt{\pi\,}} \frac{N}{\sqrt{s\,}} e^{-(u+v)^2N^2/4s}\, 
\vartheta_3\left(\II\frac{u+v}{2}\frac{N^2}{s},e^{-N^2/s}\right)
\bigl( 1 + \ldots \bigr)
\nonumber \\
&=& \frac{1}{\sqrt{\pi\,}} \frac{N}{\sqrt{s\,}} \left( \frac{N}{\sqrt{s\,}}\right)^{-2} 
\!\int_0^{N/\sqrt{s}}\!\!\!\D\alpha\int_0^{N/\sqrt{s}}\!\!\!\D\beta\: 
e^{-(\alpha+\beta)^2/4}\, \vartheta_3\left(\II\frac{\alpha+\beta}{2}\frac{N}{\sqrt{s\,}}, e^{-N^2/s}\right)
\nonumber \\
&\stackrel{N\gg\sqrt{s}}{\simeq}& \frac{1}{\sqrt{\pi\,}} \frac{s^{1/2}}{N} 
\int_0^{\infty} \!\D\alpha\int_0^{\infty}\!\D\beta\: e^{-(\alpha+\beta)^2/4}\, \bigl( 2 + \ldots \bigr)
\nonumber \\
&=& \frac{4}{\sqrt{\pi\,}}  \frac{s^{1/2}}{N} 
\EEA
where we changed the integration variables and then see that in the theta function $\vartheta_3$, 
only the leading terms are not strongly suppressed for $N^2\gg s$. 
Because of the exponential suppression in the integrand, the integration limits can be extended to infinity 
without changing much the value of the integral which rapidly converges to $2$. 
Then we have indeed the finite-size plateau height, for $s\ll N^2$ 
\BEQ \label{3.13}
C_{\infty}^{(2)} = \lim_{y\to\infty} C\bigl(s(y-1),s;0;N\bigr) = \frac{4}{\sqrt{\pi\,}}  \frac{s^{1/2}}{N} 
\EEQ
Since we had seen before that $z=2$ and $\lambda=1$, we confirm the generic expectations (\ref{1.4}). That was the main aim of this work. 
Both scaling laws quoted in (\ref{1.4}) are tested in figure~\ref{fig5}, respectively. 
Figure~\ref{fig5}a shows that $C_{\infty}^{(2)}$ decreases with $N$ while figure~\ref{fig5}b shows that $C_{\infty}^{(2)}$ increases with $s$, 
as long as $s\ll N^2$. 
Figure~\ref{fig5} also serves as an illustration 
that the asymptotic form (\ref{3.13}) already holds to good approximation for relatively small values of $s$ and $N$. 
After the kinetic spherical model \cite{Henk23}, this is only the second exactly solvable example where this kind of test was carried out exactly. 

\section{Conclusions}

We evaluated the scaled single-time and two-time correlators in the $1D$ Glauber-Ising model, quenched to $T=0$, 
on a finite-size and periodic chain. This allows to check for
the validity of the generically expected finite-size scaling forms \cite{Suzuki77} and in particular to confirm the novel expectation 
(\ref{1.4}) of the plateau height $C_{\infty}^{(2)}$. 
Both scaling laws quoted in (\ref{1.4}) are exemplified in figure~\ref{fig5}a and ~\ref{fig5}b, respectively. 
We notice that the precise form of the finite-size auto-correlator of the $1D$ Glauber-Ising model, as shown in figure~\ref{fig5}, does not
reproduce all of the generic expectations in coarsening systems, as schematised in figure~\ref{fig1}. 

In view of the general derivation \cite{Henk25}, we expect the result (\ref{1.4}) 
to hold in general, but of course further tests, including non-exactly-solvable models,
are welcome. One forthcoming case is the detailed test of (\ref{1.4}) in the $2D$ Glauber-Ising model at $T<T_c$ \cite{Wark25}. Since the exact solution of the 
$1D$ Glauber-Ising model is restricted to nearest-neighbour interactions, any tests of (\ref{1.4}) in the currently intensively studied long-range coarsening 
\cite{Corb19a,Corb19b,Corb21,Corb24,Agra21,Agra22,Agra23,Chris17,Chris19,Chris20,Chris21,Maju20,Jank23,Gess24} 
or phase-separating Ising models \cite{Muel24} require 
numerical studies. 
Here, it might also be of interest to study non-integrable generalisations of the Glauber dynamics (\ref{2.2}), for example by combining the
non-conserved Glauber-type dynamics with conserved Kawasaki-type dynamics at a different temperature, which no longer satisfies detailed balance \cite{Droz89}. 
At the least, the use of (\ref{1.4}) gives a different computational tool for the determination of $\lambda$ and $\lambda/z$, although it 
still has to be seen how precise as a numerical technique it will turn out to be. 

In addition, there is not yet any test of the generalisation of (\ref{1.4}) \cite{Henk25} 
for quenches onto $T=T_c$. Also, eventual extensions to quantum systems have not yet been tried. 
\newpage


\appsection{A}{Analytical derivations: discrete case}
For the infinite system, on the discrete chain, the analytically continued single-time correlator $C_n(t)$ 
obeys, for all $n\in\mathbb{Z}$, the equation of motion
\BEQ \label{A.1}
\partial_t C_n(t) = -2 C_n(t) + \gamma \bigl( C_{n-1}(t) + C_{n+1}(t) \bigr) 
\EEQ
and without any boundary condition. Eq.~(\ref{A.1}) is solved by a standard Fourier transform, namely 
\BEQ
\wit{C}(t,k) = \sum_{n\in\mathbb{Z}} e^{-\II k n}\, C_n(t) \;\;\; , \;\;\;
C_n(t) = \frac{1}{2\pi} \int_{-\pi}^{\pi} \!\D k\: e^{+\II k n}\, \wit{C}(t,k)
\EEQ
In Fourier space, we have
\BEQ
\partial_t \wit{C}(t,k) = -2\bigl(1 -\gamma \cos k\bigr) \wit{C}(t,k) ~~\Longrightarrow~~ \wit{C}(t,k) = \wit{C}(0,k) e^{-2(1-\gamma\cos k)t}
\EEQ
This can be re-expressed in direct space, but now we must remember from (\ref{2.5}) 
that the initial conditions $C_{-n}(0)=\alpha_n - C_{n}(0)$ for the `negative' position must 
be specified in terms of the physical initial correlators $C_{n}(0)$ with $n>0$. This gives
\BEA
\lefteqn{C_n(t) = \frac{1}{2\pi} \int_{-\pi}^{\pi} \!\D k\: \e^{\II k n}\, \wit{C}(0,k)\, e^{-2(1-\gamma\cos k) t}} \nonumber \\
&=& \sum_{m\in\mathbb{Z}} C_m(0) e^{-2t} I_{n-m}(2\gamma t) \label{A.4} \\
&=& \sum_{m\geq 0} C_m(0) e^{-2t} I_{n-m}(2\gamma t) + \sum_{m>0} C_{-m}(0) e^{-2t} I_{n+m}(2\gamma t) \nonumber \\
&=& e^{-2t} I_n(2\gamma t) + \sum_{m>0} C_m(0) e^{-2t} I_{n-m}(2\gamma t)\nonumber \\
& & + \sum_{m>0}\left\{ \left[ \left(\frac{\gamma}{1+\sqrt{1-\gamma^2\,}\,}\right)^m 
    +  \left(\frac{\gamma}{1+\sqrt{1-\gamma^2\,}\,}\right)^m \right] e^{-2t} I_{n+m}(2\gamma t) 
    - C_m(0) e^{-2t} I_{n+m}(2\gamma t) 
 \right\} \nonumber \\
&=& e^{-2t} I_n(2\gamma t) +\sum_{m>0} C_m(0) e^{-2t} \biggl[ I_{n-m}(2\gamma t) - I_{n+m}(2\gamma t) \biggr] \nonumber \\
& & +\sum_{m>0} \left[ \left(\frac{\gamma}{1+\sqrt{1-\gamma^2\,}\,}\right)^m 
    +  \left(\frac{\gamma}{1+\sqrt{1-\gamma^2\,}\,}\right)^m \right] e^{-2t} I_{n+m}(2\gamma t) 
\label{A.5}
\EEA 
where in the second line we used the integral representation of the modified Bessel function $I_n(t)$ \cite{Abra65}. 
In the forth line, we re-used the ansatz (\ref{2.5}), along with (\ref{2.7}), for the initial condition at $t=0$. 
In particular, the critical zero-temperature case $\gamma=1$ reproduces \cite[eq. (2.9)]{Henk04}. 

Sometimes, it is advantageous to reformulate this result with respect to stationary (or equilibrium) properties. 
Making the ansatz $C_{n,{\rm st}}=\eta^n$, the equilibrium  
constant $\eta$ is found from the stationarity condition $\partial_t C_{n,{\rm st}}=0$ in (\ref{A.1}). This gives the condition \cite{Glau63}
$\eta^2 - \frac{2}{\gamma}\eta+1=0$. It has the two solutions $\eta_{\pm} = \frac{1}{\gamma}\left(1\pm \sqrt{1-\gamma^2\,}\,\right)$ which are related by
$\eta_+\eta_-=1$. Clearly, the stationary (equilibrium only for $n>0$) correlator is then
\BEQ
C_{n,{\rm st}} = \eta_{-}^n = \left( \frac{1 - \sqrt{1-\gamma^2\,}\,}{\gamma} \right)^n 
\EEQ
Because of the identity, derived from  \cite[(9.6.33)]{Abra65}
\BEQ
S := e^{-2t} \sum_{m\in\mathbb{Z}} \eta^m I_{n-m}(2\gamma t) = e^{-2t} \sum_{m\in\mathbb{Z}} \eta^{m-n+n} I_{n-m}(2\gamma t) 
  = e^{-2t} \eta^n e^{\frac{1}{2} 2\gamma t \left( \eta^{-1} + \eta\right)}  = \eta^n
\EEQ
(since $\eta^{-1}+\eta = \eta_{-}+\eta_{+}=\frac{2}{\gamma}$) 
one may re-write the single-time correlator as follows, starting again at (\ref{A.4})
\BEA
C_n(t) &=& \sum_{m\in\mathbb{Z}} \left( C_m(0) -\eta_{-}^m + \eta_{-}^m \right) e^{-2t} I_{n-m}(2\gamma t) \nonumber \\
&=& \eta_{-}^n + \sum_{m\in\mathbb{Z}} \left( C_m(0) - \eta_{-}^m\right) e^{-2t} I_{n-m}(2\gamma t)  \label{A.8} \\
&=& \eta_{-}^n + \underbrace{~\left( C_0(t)-\eta_{-}^0\right)~}_{=0}e^{-2t}I_n(2\gamma t) 
+\sum_{m=1}^{\infty} \left( C_m(0)-\eta_{-}^m\right) e^{-2t}\left[ I_{n-m}(2\gamma t) - I_{n+m}(2\gamma t)\right] \nonumber 
\EEA
where we also used that $C_{-m}(0)-\eta_{-}^{-m} = C_{-m}(0) -\alpha_{-m}+\alpha_{-m}-\eta_{-}^{-m} = (-1)\left( C_m(0)-\eta_{-}^m\right)$ 
because $\alpha_{-m}=\alpha_{m}$ from 
the Lemma in section~2. The physical validity is restricted to non-negative values of $n$, because of (\ref{2.5}). 
Then we obtain the result quoted in (\ref{2.9}) in the text.

\appsection{B}{Analytical derivations: continuum limit}
The calculations for the infinite system in the continuum limit are described. 

{\bf 1.} With the analytic continuation (\ref{2.11}) 
the equation of motion of the single-time correlator has become the simple diffusion equation $\partial_t C(t;x)=\partial_x^2 C(t;x)$. It is
solved by using the Fourier representation 
$C(t;x) = \frac{1}{\sqrt{2\pi\,}}\int_{\mathbb{R}}\!\D k\: e^{\II k x} \wit{C}(t;k)$ which gives 
\BEQ \label{B.1}
\partial_t \wit{C}(t;k) = - k^2 \wit{C}(t;k) ~~\Longrightarrow~~ \wit{C}(t;k) = \wit{C}(0;k) e^{-k^2 t}
\EEQ
This is re-expressed in direct space via
\BEA
C(t;x) &=& \frac{1}{2\pi} \int_{\mathbb{R}} \!\D y\: C(0;y) \int_{\mathbb{R}}\!\D k\: e^{\II k(x-y) - k^2 t} \nonumber \\
&=& \frac{1}{\sqrt{4\pi t\,}\,} \int_{-\infty}^{\infty} \!\D y\: C(0;y)\,e^{-(x-y)^2/(4t)} \nonumber \\
&=& \frac{1}{\sqrt{4\pi t\,}\,} \left[ \int_0^{\infty}  \!\D y\: C(0;y)\,e^{-(x-y)^2/(4t)} + \int_0^{\infty}  \!\D y\: C(0;-y)\,e^{-(x+y)^2/(4t)} \right] 
\nonumber \\
&=& \erfc\left( \frac{x}{2 t^{1/2}} \right) + \frac{\exp\bigl(-\frac{x^2}{4t}\bigr)}{\sqrt{\pi\,t\,}} 
    \int_0^{\infty} \!\D y\: C(0;y)\, e^{-y^2/(4t)} \sinh\left( \frac{xy}{2t}\right)
\EEA
where we use first the solution (\ref{B.1}) in Fourier space, in the second line carried out the $k$-integration, 
in third line shall insert the analytical continuation 
(\ref{2.11}) and then compute the remaining integrals, using the definition of the complementary error function $\erfc(x)$ \cite{Abra65}. 
{}From this the physical correlator is retrieved by restricting to positive values of $x$. 
This gives (\ref{2.13}) in the text.\footnote{This might as well have been
derived from (\ref{A.8}) or (\ref{A.5}) by the asymptotic expansion \cite{Abra65} of the modified Bessel function $I_n(t)$.}  

{\bf 2.} A characteristic length scale $\ell(t)$ is found as a second moment of $C(t;x)$ as follows
\BEA
\ell^2(t) &=& \frac{\int_0^{\infty} \!\D x\: x^2 C(t;x)}{\int_0^{\infty} \!\D x\: C(t;x)} 
          \:=\: 4t\, \frac{\int_0^{\infty} \!\D x\: x^2 \erfc(x)}{\int_0^{\infty} \!\D x\: \erfc(x)} = \frac{4}{3}\, t 
\EEA
for a completely disordered initial state. 
Herein, the integrals are computed with the help of the identities \cite[(7.2.7)]{Abra65} and \cite[(7.2.14)]{Abra65}. This is (\ref{2.14}) in section~2. 

{\bf 3.} The equation of motion of the two-point correlator 
$C(\tau,s;x)=\langle \sigma(\tau+s,x)\sigma(s,0)\rangle$ is again the simple diffusion equation (\ref{2.16}).
In Fourier space this means $\wit{C}(\tau,s;k) = \wit{C}(0,s;k)\exp\bigl(-\frac{1}{2} k^2\tau\bigr)$. 
{}From the single-time correlator (\ref{2.13}), we only retain
the most relevant term (or restrict to a disordered initial state) and then have
\BEA
\lefteqn{C(\tau,s;x) = \frac{1}{2\pi} \int_{\mathbb{R}}\!\D y\: \erfc\left(\frac{|y|}{2 s^{1/2}}\right) 
\int_{\mathbb{R}}\!\D k\: e^{\II k(x-y) - \frac{1}{2}\tau k^2}}
\nonumber \\
&=& \frac{1}{\sqrt{2\pi\,\tau\,}} \int_{\mathbb{R}}\!\D y\: \erfc\left(\frac{|y|}{2 s^{1/2}}\right)\, e^{-(x-y)^2/(2\tau)} \nonumber \\
&=& \frac{\exp\bigl(-\frac{x^2}{2\tau}\bigr)}{\sqrt{2\pi\,\tau\,}} 
    \left[ \int_0^{\infty}\!\D y\: \erfc\left(\frac{y}{2 s^{1/2}}\right)\, e^{-y^2/(2\tau)+xy/\tau} 
    + \int_0^{\infty}\!\D y\: \erfc\left(\frac{y}{2 s^{1/2}}\right)\, e^{-y^2/(2\tau)-xy/\tau} \right] \nonumber \\
&=& \frac{2\exp\bigl(-\frac{x^2}{2\tau}\bigr)}{\sqrt{2\pi\,\tau\,}}  \int_0^{\infty}\!\D y\:
\left( 1 -  \erf\left(\frac{y}{2 s^{1/2}}\right)\right) e^{-y^2/(2\tau)} \cosh\left( \frac{xy}{\tau}\right)
\label{B.4}
\EEA
where in the second line the $k$-integration was carried out before in the third line the contributions of positive and negative 
$y$ are separated and are re-united in a $\cosh$-function in the forth line. 
The second of the integrals in the forth line is carried out with the help of the identity \cite[(2.8.6.5)]{Prud2}
\BEQ
\int_0^{\infty} \!\D x\: e^{-bx^2} \cosh(px) \erf(c x) 
= \frac{c}{\sqrt{\pi\,}\,b}\, \Psi_1\left(1,\frac{1}{2};\frac{3}{2},\frac{1}{2};-\frac{c^2}{b},\frac{p^2}{4b}\right)
\EEQ
involving the Humbert function $\Psi_1$ (see appendix~E) and
which can be proven by series expansion of the $\cosh$- and $\erf$-functions \cite{Abra65} and by subsequent term-wise integration. 
The first integral in (\ref{B.4}) is elementary, see e.g. \cite[(2.4.15.2)]{Prud1}. Combining all this we arrive at (\ref{2.17}) in the text. 
 
\appsection{C}{Finite-size single-time correlator}

In order to find the finite-size single-time correlator $C(t;x)$, 
we need the Fourier-transformed equation of motion. From the Fourier representation (\ref{3.4}), we have
\BEQ
\partial_t \wit{C}(t;k) = \frac{1}{2N} \int_{-N}^{N} \!\D x\: \bigl( \partial_t C(t;x)\bigr) e^{-\II\pi k {x}/{N}} 
= \frac{1}{2N} \int_{-N}^{N} \!\D x\:\bigl( \partial_x^2 C(t;x)\bigr) e^{-\II\pi k {x}/{N}} 
\EEQ
This is evaluated by repeated partial integrations and taking the periodicity conditions (\ref{3.5}) into account
\BEA
\partial_t \wit{C}(t;k) &=& \frac{1}{2N} 
\underbrace{~\left( e^{-\II\pi k}\partial_x C(t;N) - e^{\II\pi k}\partial_xC(t;-N) \right)~}_{=-2\II\,\sin(\pi k)\,\partial_x C(t;N) = 0} 
+ \frac{\II\pi k}{N} \int_{-N}^N \!\D x\: e^{-\II\pi k\frac{x}{N}} \partial_x C(t;x) \nonumber \\
&=& \frac{\II \pi k}{2N^2} \underbrace{~\left( e^{-\II\pi k} C(t;N) - e^{\II\pi k}C(t;-N) \right)~}_{=-2\II\,\sin(\pi k)\, C(t;N) = 0} 
    - \frac{\pi^2 k^2}{2 N^3} \int_{-N}^N \!\D x\: e^{-\II\pi k\frac{x}{N}}  C(t;x) \nonumber \\
&=& - \left( \frac{\pi k}{N}\right)^2 \wit{C}(t;k) 
\EEA 
where in the first and second lines the periodicity conditions (\ref{3.5}) 
were used and the boundary terms vanish since $k\in\mathbb{Z}$.  The obvious solution
\BEQ
\wit{C}(t;k) = \wit{C}(0;k)\exp\left[ - \left(\frac{\pi k}{N}\right)^2 t\right]
\EEQ
is readily transformed back into real space, with the help of (\ref{2.12})
\BEA
C(t;x) &=& \sum_{k=-\infty}^{\infty} \wit{C}(0;k) \exp\left( - \frac{\pi^2 k^2 t}{N^2} + \II\pi k \frac{x}{N} \right) \nonumber \\
&=& \frac{1}{2N} \int_{-N}^N \!\D y\: C(0;y) 
\underbrace{~\sum_{k=-\infty}^{\infty}\exp\left(-\frac{\pi^2 k^2 t}{N^2} 
            + \II\pi k\frac{x-y}{N}\right)~}_{=\vartheta_3\bigl(\frac{\pi}{2}\frac{x-y}{N},e^{-\pi^2 t/N^2}\bigr)}
\nonumber \\
&=& \frac{1}{2N} \int_{0}^N \!\D y\: \left[ C(0;y) \vartheta_3\left(\frac{\pi}{2} \frac{x-y}{N},e^{-\pi^2 t/N^2}\right) 
                                       + C(0;-y) \vartheta_3\left(\frac{\pi}{2} \frac{x+y}{N},e^{-\pi^2 t/N^2}\right) \right]
\nonumber \\
&=& \frac{1}{2N} \int_{0}^N \!\D y\: \left\{ 2\vartheta_3\left(\frac{\pi}{2} \frac{x+y}{N},e^{-\pi^2 t/N^2}\right) \right. \nonumber \\
& & \left. +C(0;y)\left[\vartheta_3\left(\frac{\pi}{2}\frac{x-y}{N},e^{-\pi^2 t/N^2}\right)
           -\vartheta_3\left(\frac{\pi}{2}\frac{x+y}{N},e^{-\pi^2 t/N^2}\right)\right]\right\}
\label{C.4}
\EEA
where $\vartheta_3(z,q)$ is a Jacobi theta function \cite[(16.27.4)]{Abra65}. 
In view of what was said in section~2 about the infinite-size system, we expect that the second line
in (\ref{C.4}) is negligible for times $t\gg 1$, if only $C(0;x)\to 0$ decreases with $x$. See figure~\ref{fig4}a for an illustration. 

A convenient reformulation of this result applies the following identities, namely
\BEA
\lefteqn{\hspace{-2cm}\frac{1}{N} \int_0^{N} \!\D y\: \vartheta_3\left(\pi \frac{x+y}{2N},e^{-\pi^2 t/N^2}\right) =
\frac{1}{N} \int_0^N \!\D y\: \sum_{k=-\infty}^{\infty} \exp\left(-\frac{\pi^2 k^2 t}{N^2} + \II\pi k\frac{x+y}{N} \right)} \nonumber \\
&=& 1 +\frac{2}{\pi} \sum_{k=1}^{\infty}  e^{-\pi^2 k^2 t/N^2} \frac{1}{k} 
    \left( \sin\left(\pi k\frac{x-N}{N}\right) - \sin\left(\pi k\frac{x}{N}\right) \right) 
\nonumber \\
&=& 1 +\frac{2}{\pi} \sum_{k=1}^{\infty}  e^{-\pi^2 k^2 t/N^2} \frac{1}{k}  \sin\left(\pi k\frac{x}{N}\right)  \left( (-1)^k -1 \right) 
\nonumber \\
&=& 1 - \frac{4}{\pi} \sum_{k=0}^{\infty} e^{-\pi^2 (2k+1)^2 t/N^2}  \frac{1}{2k+1}  \sin\left(\pi (2k+1)\frac{x}{N}\right) \\
&=& 1 - \frac{2}{\pi N} \sum_{k=0}^{\infty} e^{-\pi^2 (2k+1)^2 t/N^2} \int_0^x \!\D x' \cos\left( \pi (2k+1)\frac{x'}{N}\right) \nonumber \\
&=& 1 - \frac{2}{\pi} \int_0^{x/N} \!\D v\: \vartheta_2\left(\pi v, e^{-\bigl(\frac{2\pi}{N}\bigr)^2 t} \right)
\EEA
where first the auxiliary integration over $y$ is carried out and then we see that only the odd values of $k$ contribute before comparing with the other
Jacobi theta function $\vartheta_2(z,q)$ \cite[(16.27.2)]{Abra65}; and then 
\BEA
& & \frac{1}{2N} \int_{0}^N \!\D y\:
C(0;y) \left[ \vartheta_3\left(\pi \frac{x-y}{2N},e^{-\pi^2 t/N^2}\right) - \vartheta_3\left(\pi \frac{x+y}{2N},e^{-\pi^2 t/N^2}\right) \right]
\nonumber \\
&=&  \frac{1}{2N} \int_{0}^N \!\D y\: C(0;y) \sum_{k\in\mathbb{Z}} e^{-\pi^2 k^2 t/N^2}
\biggl( e^{\II \pi k (x-y)/N} - e^{-\II \pi k (x-y)/N} \biggr) 
\nonumber \\
&=& \frac{1}{N} \sum_{k=1}^{\infty} e^{-\pi^2 k^2 t/N^2}  \int_{0}^N \!\D y\: C(0;y)  
    \left( \cos\left(\pi k\frac{x-y}{N}\right) - \cos\left(\pi k\frac{x+y}{N}\right) \right) 
\nonumber \\
&=& \frac{2}{N}  \sum_{k=1}^{\infty} e^{-\pi^2 k^2 t/N^2}  \sin\left(\pi k\frac{x}{N}\right) 
    \int_{0}^N \!\D y\: C(0;y) \sin\left(\pi k\frac{y}{N}\right)  
\EEA
which follows from trigonometric addition theorems. 
Using these identities in (\ref{C.4}), 
we find
\BEA
C(t;x) &=& 1 - \frac{4}{\pi} \sum_{k=0}^{\infty} \frac{e^{-\pi^2(2k+1)^2 t/N^2}}{2k+1} \sin\frac{\pi(2k+1)x}{N} \nonumber \\
& & + \frac{2}{N} \sum_{k=1}^{\infty} e^{-\pi^2 k^2 t/N^2}\sin\frac{\pi kx}{N} \int_0^N\!\D y\: C(0;y)  \sin\frac{\pi ky}{N} 
\label{C.7}
\EEA
for any initial correlator $C(0;x)$. Together, this gives (\ref{3.7}) in the text. 
Clearly, these representations are periodic in $x$, with period $2N$. In addition, we also have from (\ref{C.7}) that 
\BEA
C(t;N-x) &=& 1 - \frac{4}{\pi} \sum_{k=0}^{\infty} \frac{e^{-\pi^2(2k+1)^2 t/N^2}}{2k+1} \sin\frac{\pi(2k+1)(N-x)}{N} \nonumber \\
& & + \frac{2}{N} \sum_{k=1}^{\infty} e^{-\pi^2 k^2 t/N^2}\sin\frac{\pi k(N-x)}{N} \int_0^N\!\D y\: C(0;y)  \sin\frac{\pi ky}{N} \nonumber \\
&=& 1 - \frac{4}{\pi} \sum_{k=0}^{\infty} \frac{e^{-\pi^2(2k+1)^2 t/N^2}}{2k+1} \cos\bigl(\pi(2k+1)\bigr)\sin\frac{\pi(2k+1)(-x)}{N} \nonumber \\
& & + \frac{2}{N} \sum_{k=1}^{\infty} e^{-\pi^2 k^2 t/N^2}\cos\bigl({\pi k }\bigr)\sin\frac{\pi k(-x)}{N} \int_0^N\!\D y\: C(0;N-y)  \sin\frac{\pi k(N-y)}{N} 
\nonumber \\
&=& 1 - \frac{4}{\pi} \sum_{k=0}^{\infty} \frac{e^{-\pi^2(2k+1)^2 t/N^2}}{2k+1} \big(-1\bigr)^2\sin\frac{\pi(2k+1)x}{N} \nonumber \\
& & + \frac{2}{N} \sum_{k=1}^{\infty} e^{-\pi^2 k^2 t/N^2}\big( (-1)^k\bigr)^2 \big(-1\bigr)^2\sin\frac{\pi kx}{N} 
    \int_0^N\!\D y\: C(0;y)  \sin\frac{\pi k y}{N} 
\nonumber \\
&=& C(t;x)
\EEA
which proves the required inversion relation. In this, we changed in the second line the integration variable, applied the trigonometric addition theorem 
and used in the third line that we have the inversion relation $C(0;N-x)=C(0;x)$ for the initial configuration. 

Finally, the Jacobi theta function $\vartheta_3\bigl(\frac{\pi}{2}u,q\bigr) = \vartheta_3\bigl(\frac{\pi}{2}(2-u),q\bigr)$ 
obeys the following modular transformation identity, to be shown via Poisson's resummation formula  \cite{Itzy89}
\BEQ \label{C.9}
\vartheta_3\left( \pi u, e^{-\pi t}\right) 
= t^{-1/2} \exp\left(-\pi \frac{u^2}{t} \right) \vartheta_3\left( \II\pi \frac{u}{t},e^{-\pi/t}\right)
\EEQ

For the proof of (\ref{3.8}), we need a few preparations. First, with the change of variables $w=u+v$ and $w'=u-v$ and $f(u,v)=g(w,w')$, 
the square domain of integration $[0,1]^2$ is decomposed into two triangles and we have
\BEA
\int_0^1\!\D u\int_0^1\!\D v\: f(u,v) &=& 
\frac{1}{2} \int_0^1 \!\D w\int_{-w}^{w} \!\D w'\: g(w,w') + \frac{1}{2} \int_1^2 \!\D w\int_{-(2-w)}^{2-w} \!\D w'\: g(w,w')
\nonumber \\
&=& \frac{1}{2} \int_0^1 \!\D w\int_{-w}^{w} \!\D w'\:  \bigl( g(w,w') + g(2-w,w') \bigr)
\label{C.11} 
\EEA
Then we have, with (\ref{C.4}) for a disordered initial state, $q=e^{-\pi\bigl(\pi t/N^2\bigr)}$ and the properties of $\vartheta_3$ 
\BEA
\lefteqn{ \frac{1}{N} \int_0^N \!\D x\: C(t;x) = \int_0^1 \D v \int_0^1 \!\D u\: \vartheta_3\left( \frac{\pi}{2}(v+u), q\right)} \nonumber \\
&=& \frac{1}{2} \int_0^1 \!\D w\int_{-w}^{w} \!\D w'\: \left[ \vartheta_3\left(\frac{\pi}{2} w, q\right) 
    + \vartheta_3\left(\frac{\pi}{2} (2-w), q\right) \right] 
= 2 \int_0^1 \!\D w\: w\, \vartheta_3\left(\frac{\pi}{2} w, q\right)~~~~
\EEA
and, again using (\ref{C.11}) 
\BEA
\lefteqn{ \frac{1}{N} \int_0^N \!\D x\: \bigl(\frac{N}{\pi} \sin(\frac{\pi x}{N})\bigr)^2 C(t;x) 
= \frac{N^2}{\pi^2} \int_0^1 \D v \int_0^1 \!\D u\: \sin^2(\pi v)\, \vartheta_3\left( \frac{\pi}{2}(v+u), q\right)~~~~} \nonumber \\
&=& \frac{N^2}{4\pi^2}\int_0^1 \!\D w\: \vartheta_3\left(\frac{\pi}{2} w, q\right) \int_{-w}^{w} \!\D w'\: \bigl( 1 - \cos(\pi(w-w')\bigr)  \nonumber \\
&=& \frac{N^2}{4\pi^2}\int_0^1\!\D w\: \vartheta_3\left(\frac{\pi}{2} w, q\right) \bigl( 2w - \frac{1}{\pi} \sin 2\pi w \bigr)
\EEA
The quotient of these gives the integral form for $\ell(t)$ in (\ref{2.14}). Deep into the finite-size regime, one has $t\gg N^2$, so that $\vartheta_3\to 1$. 
This establishes the first asymptotic form. For the second one, we use the modular identity (\ref{C.9}) and find 
\BEQ
\ell^2(t) = \frac{2 N^2}{\pi^2} \left[ 1 - \frac{1}{2\pi}
\frac{\int_0^1 \!\D w\: \sin(2\pi w)\, e^{-N^2 w^2/t}\, 
\vartheta_3\bigl(\II \pi N^2 w/(2t),e^{-N^2/t}\bigr)}{\int_0^1 \!\D w\: w\, e^{-N^2 w^2/t}\, 
\vartheta_3\bigl(\II \pi N^2 w/(2t),e^{-N^2/t}\bigr)} \right]
\EEQ
For early times and large systems, $t\ll N^2$ such that now $e^{-N^2/t}\to 0$ and again $\vartheta_3\to 1$. Then 
\BEA
\ell^2(t) &\simeq& \frac{2 N^2}{\pi^2} \left[ 1 - 
\frac{\int_0^1 \!\D w\: \sin(2\pi w) e^{-N^2 w^2/t} \bigl( 1 + \ldots \bigr)}{\int_0^1 \!\D w\: 2\pi w\, e^{-N^2 w^2/t} \bigl( 1 + \ldots \bigr)} \right] 
\nonumber \\
&\simeq& \frac{2 N^2}{\pi^2} \left[ 1 - 1 + \frac{\int_0^1 \D z\: \frac{4\pi^2}{6} z e^{-N^2 z/t}}{\int_0^1 \D z\:  e^{-N^2 z/t}} \right]
\left( 1 +{\rm o}(t N^{-2})\right) \nonumber\\
&\simeq&  \frac{4}{3}\, t \left( 1 +{\rm o}(t N^{-2})\right)
\EEA
reproduces (\ref{2.14}), as claimed. 

\appsection{D}{Finite-size two-time correlator}

As a first step, we must derive the Fourier-transformed equation of motion (\ref{3.9}). 
Because of the periodicity (\ref{3.5}) of the single-time correlator
$C(0,s;x)=C(s;x)$, one can integrate forward in time to establish that
\BEQ
C(\tau,s;N) = C(\tau,s;-N) \;\; , \;\; \partial_x C(\tau,s;N) = \partial_x C(\tau,s;-N)
\EEQ
and the analytically continued two-time correlator is a periodic function in $x$ of period $2N$. With the Fourier representation (\ref{3.4}), 
it is immediate to repeat the steps of appendix~C to arrive at the equation
\BEQ \label{D.2}
\partial_{\tau} \wit{C}(\tau,s;k) = -\frac{1}{2} \left(\frac{\pi k}{N}\right)^2 \wit{C}(\tau,s;k)
\EEQ
The obvious solution is then re-transformed into direct space, in analogy with (\ref{B.4}) in appendix~B
\BEA
\lefteqn{C(\tau,s;x) = \sum_{k=-\infty}^{\infty} \wit{C}(0,s;k)\, e^{-\frac{1}{2}\bigl(\frac{\pi k}{N}\bigr)^2 \tau}} 
\nonumber \\
&=&\frac{1}{2N}\sum_{k=-\infty}^{\infty} \int_0^N \!\D y\: C(0,s;y) 
   \left( e^{\II\pi k(x-y)/N} + e^{\II\pi k(x+y)/N}\right) e^{-\frac{1}{2}\bigl(\frac{\pi k}{N}\bigr)^2 \tau} 
\nonumber \\
&=& \frac{1}{2N}\int_0^N \!\D y\int_0^1 \!\D u\: \vartheta_3\left(\frac{\pi}{2}\frac{y}{N}+\frac{\pi}{2}u,e^{-\pi^2 s/N^2}\right) \times 
\nonumber \\
& & \times \left[  \vartheta_3\left(\frac{\pi}{2}\frac{x-y}{N},e^{-\frac{1}{2}\frac{\pi^2}{N}\tau}\right)
                 + \vartheta_3\left(\frac{\pi}{2}\frac{x+y}{N},e^{-\frac{1}{2}\frac{\pi^2}{N}\tau}\right) \right] 
\\
&=& \frac{1}{2}\int_0^1 \!\!\D v\int_0^1 \!\!\D u\: \vartheta_3\left(\frac{\pi}{2}(v+u),e^{-\frac{\pi^2 s}{N^2}}\right)
\left[ \vartheta_3\left(\frac{\pi}{2}\frac{x}{N}-\frac{\pi}{2}v,e^{-\frac{\pi^2}{N^2}\frac{\tau}{2}}\right)
     + \vartheta_3\left(\frac{\pi}{2}\frac{x}{N}+\frac{\pi}{2}v,e^{-\frac{\pi^2}{N^2}\frac{\tau}{2}}\right) \right] 
\nonumber 
\EEA
which is (\ref{3.10}) in section~3. Herein, we used in the second line the solution of the equation of motion (\ref{D.2}), 
then inserted the relevant part of
the single-time correlator (\ref{C.4}) in the third line and then re-scaled the integration variable. 
Repeating the argument from appendix~C, we also have $C(\tau,s;N-x)=C(\tau,s;x)$. 

\appsection{E}{On Humbert function identities}
The Humbert function $\Psi_1$ is defined \cite{Sriv85,Hang24a,Hang24b,Hang25} in terms of a power series (converging for $|x|<1$ and $|y|<\infty$), 
to which we add an useful integral representation \cite{Wald18} for the analytical continuation to all negative arguments
\begin{subequations}
\begin{align}
\Psi_1(a,b;c,c';x,y) &=
\sum_{n=0}^{\infty} \sum_{m=0}^{\infty} \frac{(a)_{n+m} (b)_n}{(c)_n (c')_m} \frac{x^n}{n!}\frac{y^m}{m!} \label{E.1a} \\
&= \frac{1}{\Gamma(a)} \int_0^{\infty} \!\D u\: e^{-u} u^{a-1}\: {}_1F_{1}(b;c;xu)\: {}_0F_{1}(c';yu)     \label{E.1b}
\end{align}
\end{subequations}
with the Pochhammer symbol $(a)_n$ and the generalised hyper-geometric functions ${}_1F_1$ and ${}_0F_1$ \cite{Prud3}. 
In view of (\ref{2.17}), several asymptotic identities on the Humbert function $\Psi_1\left(1,\frac{1}{2};\frac{3}{2},\frac{1}{2};x,y\right)$ 
are stated, which are needed in the text. 
These asymptotic relations are special cases of more general theorems proven in \cite{Hang25b}. 
It is important that the second argument in the results which follow is always positive.

\noindent
{\bf Lemma E.1.} {\it One has the identity, where $\xi>0$ is kept fixed}
\BEQ
\lim_{z\to 0^+} e^{-\xi^2/2z} z^{1/2} \Psi_1\left(1,\frac{1}{2}; \frac{3}{2},\frac{1}{2};-z, \frac{\xi^2}{2z} \right) = \frac{\pi}{2} \erf( \xi/\sqrt{2} )
\EEQ
This Lemma expresses mathematically the reduction from the two-time correlator to the single-time one, in the equal-time limit $\tau=t-s\to 0$. \\

\noindent {\bf Proof:} This straightforward calculation starts from the integral representation (\ref{E.1b}). Then  
\BEA
\lefteqn{ 
\Psi_1\left(1,\frac{1}{2}; \frac{3}{2},\frac{1}{2};-z, \frac{y^2}{z} \right) 
= \int_0^{\infty} \!\D u\: e^{-u}\, {}_1F_1\left(\frac{1}{2};\frac{3}{2};-zu\right) {}_0F_{1}\left(\frac{1}{2}; \frac{y^2}{z} u\right) 
} \nonumber \\
&=& \frac{1}{z} \int_0^{\infty} \!\D v\: e^{-v/z}\, {}_1F_1\left(\frac{1}{2};\frac{3}{2};-v\right) \cosh\left( 2 \sqrt{ \frac{y^2}{z^2} v\,}\, \right) 
\nonumber \\
&=& \frac{1}{z} \int_0^{\infty} \!\D w\: w\,  {}_1F_1\left(\frac{1}{2};\frac{3}{2};-w^2\right)
    \left( e^{-\frac{1}{z}[ (w-y)^2-y^2]} + e^{-\frac{1}{z}[ (w+y)^2-y^2]} \right)
\nonumber 
\EEA
where in the second line we used the identity \cite[(7.13.1.5)]{Prud3} and tacitly admit $y>0$. 
When $z\to 0^+$, the first term will have its main contribution from $w\approx y>0$ and the second one from 
$w\approx -y<0$, but this will be out of the integration domain. 
We can then write
\BD
e^{-y^2/z} z^{1/2} \Psi_1\left(1,\frac{1}{2}; \frac{3}{2},\frac{1}{2};-z, \frac{y^2}{2z} \right) =
\int_0^{\infty} \!\D w\: w\,  {}_1F_1\left(\frac{1}{2};\frac{3}{2};-w^2\right)
    \frac{\sqrt{\pi\,}}{\sqrt{\pi z\,}\,} \left( e^{-\frac{1}{z}(w-y)^2} + e^{-\frac{1}{z}(w+y)^2} \right)
\ED
For $z\to 0^+$, we recognise that the first term in the bracket will give a representation of the Dirac delta function
\BEQ \label{E.3} 
\delta(w) = \lim_{z\to 0^+} \frac{1}{\sqrt{\pi\, z\,}\,}\, e^{- w^2/z}
\EEQ
so that we finally have
\BEA
& & \hspace{-1cm}\lim_{z\to 0^+} e^{-y^2/z} z^{\frac{1}{2}}\, \Psi_1\left(1,\frac{1}{2}; \frac{3}{2},\frac{1}{2};-z, \frac{y^2}{z} \right) 
\:=\: \int_0^{\infty} \!\!\!\D w\: w\,  {}_1F_1\left(\frac{1}{2};\frac{3}{2};-w^2\right) \sqrt{\pi\,}\, \bigl( \delta(w-y) + \delta(w+y) \bigr) 
\nonumber \\
&=& \sqrt{\pi\,}\, y\: {}_1F_1\left( \frac{1}{2};\frac{3}{2};-y^2\right) 
\:=\: \frac{\pi}{2} \erf(y)
\nonumber
\EEA
where in the last step the identities \cite[(7.11.3.1)]{Prud3} and \cite[(6.5.17)]{Abra65} were used.  \hfill q.e.d. 

\noindent {\bf Lemma E.2.} \cite[Theorem 3.6]{Hang25b} {\it In the limit where $x\to+\infty$ and $y\in\mathbb{R}$ is kept fixed, one has}
\BEQ \label{E.4}
\Psi_1\left(1,\frac{1}{2};\frac{3}{2},\frac{1}{2};-x,\frac{y^2}{x}\right) 
\simeq \frac{\pi}{2}\frac{1}{x^{1/2}} - \frac{1}{x} + \frac{y^2}{x^{3/2}} + \frac{1}{3} \frac{1}{x^2} + \ldots
\EEQ
The two leading terms do not depend on $y$. This is needed for deriving (\ref{2.19}). \\

\noindent {\bf Lemma E.3.} {\it In the limit where $y\to\infty$ and $x>0$ is kept fixed, one has to leading order in $1/y$}
\BEQ \label{E.5}
\Psi_1\left(1,\frac{1}{2};\frac{3}{2},\frac{1}{2};-x,\frac{y^2}{x}\right) 
\simeq \frac{\pi}{2} \frac{1}{x^{1/2}}\, e^{y^2/x} - \frac{\sqrt{\pi\,}}{2} \frac{1}{y}\frac{1+x}{x^{1/2}}\, e^{y^2/[(1+x)x]}
\EEQ
We notice the presence of two distinct exponential terms, both of which are needed in the text, for the derivation of (\ref{2.20}). 

\noindent
{\bf Heuristic argument.} {We start from the integral representation (\ref{E.1b}) \cite{Wald18}. It turns out that the contributions of 
interest to us come from the upper limit of the integration whereas the lower limit 
merely contributes terms of algebraic size which can be discarded. So we may split
the relevant integral $\int_0^{\infty} = \int_0^{\eta x/y^2} + \int_{\eta x/y^2}^{\infty}$ where $\eta$ is some constant to be fixed later. 
We concentrate on the second term which reads
\BEA
\mathfrak{T}_2^{(\eta)} &=& \frac{1}{\Gamma(a)} \int_{\eta x/y^2}^{\infty} \!\D u\: e^{-u} u^{a-1} {}_1F_{1}(b;c;-xu) {}_0F_{1}(c'; y^2 u/x) 
\nonumber \\
&=& \frac{2\Gamma(c')}{\Gamma(a)} y^{1-c'} x^{c'/2-1/2} \int_{(\eta x)^{1/2}/y}^{\infty} \!\D w\: e^{-w^2} w^{2a-c'} {}_1F_{1}(b;c;-x w^2) 
    I_{c'-1}\left(2 \frac{y}{x^{1/2}} w\right) 
\nonumber 
\EEA
where \cite[(7.13.1.1)]{Prud3} was used. Since $y$ will be large, we need the large-argument expansions of ${}_1F_1$ and $I_{c'-1}$, 
eqs. (13.5.1) and (9.7.1) of \cite{Abra65}, of which we shall only write the first term, respectively. Formally, 
\BEA
\mathfrak{T}_2^{(\eta)} &\simeq& 
\frac{2\Gamma(c')}{\Gamma(a)} y^{1-c'} x^{c'/2-1/2} \int_{(\eta x)^{1/2}/y}^{\infty} \!\D w\: e^{-w^2} w^{2a-c'} \times \nonumber \\
& & \times \left\{ \frac{\Gamma(c)}{\Gamma(c-b)} \bigl( x w^2\bigr)^{-b} \bigl( 1 + \ldots\bigr) 
+ \frac{\Gamma(c)}{\Gamma(b)} e^{-x w^2} \bigl( -x w^2\bigr)^{b-c} \bigl( 1 + \ldots\bigr) \right\} 
\cdot \frac{\exp\left[ 2 y\, x^{-1/2} w\right]}{\sqrt{2\pi \cdot 2 y\, x^{-1/2} w\,}\,} \bigl( 1 + \ldots\bigr) 
\nonumber \\
&\simeq& \frac{\Gamma(c)\Gamma(c')}{\Gamma(a)\Gamma(c-b)} y^{-1/2-c'} x^{c'/2-b-1/4}
    \int_0^{\infty} \!\D w\: w^{2a-2b-c'-1/2} e^{-[(w-y x^{-1/2})^2 - y^2/x]} \frac{y}{\sqrt{\pi\,}\,} \nonumber \\
& & + \frac{\Gamma(c)\Gamma(c')}{\Gamma(a)\Gamma(b)} y^{1/2-c'} x^{b-c+c'/2-1/4} 
      \int_0^{\infty} \!\D w\:  w^{2a+2b-2c-c'-1/2}   \times \nonumber \\
& & \times \exp\left( - \left[ \left(1+x)^{1/2} w - \frac{y}{[(1+x)x]^{1/2}}\right)^2 - \frac{y^2}{(1+x)x}\right] \right)
      \frac{y}{\sqrt{\pi\,}\,}  e^{\II\pi(b-c)} 
\nonumber \\
&=& \frac{\Gamma(c)\Gamma(c')}{\Gamma(a)\Gamma(c-b)} y^{-1/2-c'} x^{c'/2-b-1/4}
    \int_0^{\infty} \!\D v\: y^{2a-2b-c'+1/2} v^{2a-2b-c'-1/2} 
    \textcolor{blau}{\underbrace{~e^{-y^2\left(v-x^{-1/2}\right)^2}~}_{\to \delta(v-x^{-1/2}) \sqrt{\pi\,}/y}} \: e^{y^2/x} \frac{y}{\sqrt{\pi\,}\,} 
    \nonumber \\  
& & + \frac{\Gamma(c)\Gamma(c')}{\Gamma(a)\Gamma(b)} y^{1/2-c'} x^{b-c+c'/2-1/4} 
      \int_0^{\infty} \!\D v\:  y^{2a+2b-2c-c'+1/2} v^{2a+2b-2c-c'-1/2}  
      \times \nonumber \\
&& \times (1+x)^{-1/4-a-b+c+c'/2} \textcolor{blau}{\underbrace{~e^{-y^2\left( v - [(1+x)x]^{-1/2}\right)^2}~}_{\to \delta(v-[(1+x)x]^{-1/2})\sqrt{\pi\,}/y}} 
   \: e^{y^2/[(1+x)x]} \frac{y}{\sqrt{\pi\,}\,}  e^{\II\pi(b-c)} 
\nonumber \\
&\simeq& \frac{\Gamma(c)\Gamma(c')}{\Gamma(a)\Gamma(c-b)} e^{y^2/x} \left( \frac{y^2}{x}\right)^{a-b-c'} x^{-b} \nonumber \\
& & + \frac{\Gamma(c)\Gamma(c')}{\Gamma(a)\Gamma(b)} e^{y^2/[(1+x)x]} \left(\frac{y^2}{x}\right)^{a+b-c-c'} \bigl( -x\bigr)^{b-c} 
(1+x)^{-2a-2b+2c+c'+1/2} 
\nonumber
\EEA
and specialising to $a=1$, $b=\frac{1}{2}$, $c=\frac{3}{2}$ and $c'=\frac{1}{2}$ gives the assertion. 
In the second line, the integrals were extended down to zero, admissible for $y\gg 1$ 
and in the third line, after a change of variables, we recognise the Dirac delta function (\ref{E.3}) for $y\gg 1$.\\
This is merely a heuristic argument. In principle, both terms in (\ref{E.5}) should be the first ones of an asymptotic series. 
It remains an open mathematical problem why there should be no such series associated with the first term. 
Mathematically, (\ref{E.5}) remains a conjecture.

For completeness, we add the following. It is well-known \cite[Theorem 3.6]{Hang24b}, \cite[Theorem 3.4 and 3.6]{Hang25}
that if $x>0$ and $y>0$ are fixed constants, and with the necessary conditions on $a,b,c,c'$ one has
\begin{align} 
\lim_{t\to\infty} \biggl( e^{-t y} t^{2b+c'-a}\, \Psi_1(a,b;c,c';-tx,ty)\biggr) &= \frac{\Gamma(c)\Gamma(c')}{\Gamma(a)\Gamma(c-b)} \,
\left(\frac{y}{x}\right)^{b} {y^{a-2b-c'}}
\label{E.6b}
\end{align}
This result also leads to the kind of singularity of $\Psi_1$ when $x\to 1^{-}$ from the left, namely 

\noindent{\bf Lemma E.4.} {\it If $y>0$ and with the necessary conditions on $a,b,c,c'$, we have}
\begin{align} 
& \lim_{x\to 1^{-}} \biggl( e^{-y/(1-x)} (1-x)^{2a+2b-2c-c'}\, \Psi_1(a,b;c,c';x,y)\biggr) = 
\frac{\Gamma(c)\Gamma(c')}{\Gamma(a)\Gamma(b)} \,{y^{a+b-c-c'}}
\label{E.7b}
\end{align}

\noindent{\bf Proof.} {For $x<1$, this is based on the Kummer-type relation \cite[(2.8)]{Hang25}
\BEQ \label{E.8}
\Psi_1(a,b;c,c';x,y) = (1-x)^{-a} \Psi_1\left( a,c-b;c,c';-\frac{x}{1-x}, \frac{y}{1-x}\right)
\EEQ
In view of (\ref{E.8}), define a new variable $t$ and observe that when $x\to 1^{-}$ from the left
\BD
t := \frac{1}{1-x} \to \infty \;\; , \;\; \frac{x}{1-x} = t\left( 1 - \frac{1}{t}\right) = t-1 \to \infty \tag{*}
\ED
so that we can write, with $y>0$  being kept fixed,
\BEA
\lim_{x\to 1^{-}} e^{-y/(1-x)} (1-x)^{2a-2(c-b)-c'}\,\Psi_1(a,b;c,c';x, +y) &=& \lim_{t\to\infty} e^{-ty} t^{2(c-b)+c'-a} \Psi_1(a,c-b;c,c';-t, +ty) \nonumber
\EEA
Using (*), we recast (\ref{E.6b}) into the form presently required.} \hfill q.e.d. 

As an example, as $x\to 1^{-}$ and $y>0$, read off: 
$\Psi_1\left(1,\frac{1}{2};\frac{3}{2},\frac{1}{2};x,y\right)\approx \sqrt{\frac{\pi}{4y}\,}\cdot\bigl(1-x\bigr)^{\frac{1}{2}}\,e^{y/(1-x)}$.


\noindent
{\bf Acknowledgements:} It is a pleasure to thank D. Warkotsch, W. Janke, P.-C. Hang for interesting discussions and correspondence. 
This work was supported by the french ANR-PRME UNIOPEN (ANR-22-CE30-0004-01).  


\end{document}